\documentclass[12pt,english]{article}
\usepackage{booktabs,caption}
\usepackage[flushleft]{threeparttable}
\usepackage{makeidx}
\usepackage{amssymb,bm}
\usepackage{amsmath}
\usepackage{mathrsfs}
\usepackage[scr=rsfs,cal=boondox]{mathalfa}
\usepackage{amsthm}
\usepackage{amssymb}

\usepackage{float}
\restylefloat{table}
\usepackage{fancyhdr}
\usepackage[comma,authoryear]{natbib}
\usepackage{graphicx}

\fancyheadoffset{0\textwidth}
\fancypagestyle{plain}{%
	\fancyhf{} 
	\fancyhead{}
	\fancyfoot{}
	\fancyhead[R]{\thepage} 
	
	}
\newcommand{\independent}{\rotatebox[origin=c]{90}{$\models$}}
\usepackage{mathtools}
\usepackage{setspace}

\usepackage[bookmarks]{hyperref}
\usepackage{comment}
\hypersetup{
	bookmarks=true,         
	unicode=false,          
	pdftoolbar=true,        
	pdfmenubar=true,        
	pdffitwindow=true,      
	pdftitle={Causal Overlap Effects: A Cumulative Fixed Effect Approach},
	pdfauthor={Jingying He and Felix Elwert},
	pdfsubject={Subject},   
	pdfnewwindow=true,      
	pdfkeywords={keywords}, 
	colorlinks=true,        
	linkcolor=black,         
	citecolor=black,        
	filecolor=magenta,      
	urlcolor=black           
}
\usepackage[margin=1.25in]{geometry}
\usepackage{scalefnt}
\usepackage{pifont}
\usepackage{tikz}
\usetikzlibrary{shapes,arrows}

\usepackage{pdflscape}

\usepackage{endnotes}
\usepackage{multirow}

\let\footnote=\endnote

\makeatletter

\newcommand\backmatter{%
	\if@openright
	\cleardoublepage
	\else
	\clearpage
	\fi
}

\usepackage{array}
\newcolumntype{P}[1]{>{\centering\arraybackslash}p{#1}}
\newcolumntype{M}[1]{>{\centering\arraybackslash}m{#1}}

\makeatother

\setcitestyle{citesep={;}}

\begin{document}

\begin{titlepage}
\title{Causal Overlap Effects: A Cumulative Fixed Effect Approach\thanks{The paper benefitted from comments by Xi Song, Eric Grodsky, Ted Gerber, and Junlong Feng. Previous versions presented at PAA. This research was conducted at the Center for Demography and Ecology and the Center for the Demography of Health and Aging at the University of Wisconsin-Madison (P2C HD047873).}
}
\author{Jingying He\thanks{jingyinghe@gmail.com} \and  Felix Elwert\thanks{Department of Sociology and Center for Demography and Ecology, University of Wisconsin-Madison; elwert@wisc.edu} }
\maketitle
\begin{abstract}
\noindent
Sociologists often ask about the effect of increasing one's duration of exposure to a social context on one's outcomes, i.e. the overlap effect. Past studies adopted a unidimensional treatment effect framework to estimate the effect of overlap, imposing important restrictions. In this paper, we propose a new causal framework of multidimensional treatments where the overlap effects include both the duration and the content of overlap, under which, for instance, the grandparent overlap effect is defined as the union of all causal effects of a grandparent's observed and unobserved characteristics (i.e., the content) on the grandchild across their shared life course (i.e, the duration). The multidimensional framework allows for a more flexible and sociologically rich approach to effect heterogeneity, where unobserved contextual characteristics play two roles- as unobserved confounders and as integral components of overlap effects- overlap effects in this framework are not easily estimated with conventional fixed effects estimation. Hence, we develop a new cumulative fixed effects (CFE) approach that can estimate a range of interesting heterogeneous causal overlap effects from three-wave individual panel data. We show that the CFE approach is unbiased even in highly non-linear simulations, and we discuss assumptions and extensions.

\noindent\textbf{Keywords:} overlap time, fixed effects, causal inference.  \\
\bigskip
\end{abstract}
\setcounter{page}{0}
\thispagestyle{empty}
\end{titlepage}

\doublespacing

\section{Introduction}
Research in sociology pertains to the effect of overlapping time to a social context on one's social-economic outcomes. For instance, family sociologists are concerned about the duration of fathers' presence on children's development (\citealt{magnuson2009family,mclanahan2013causal}). Educational sociologists are interested in the effect of increasing teacher's tenure on student achievements (\citealt{simon2015teacher,ingersoll2001teacher}). Neighborhood researchers investigate the duration of exposure to an impoverished neighborhood on teenagers' behaviors (\citealt{timberlake2007racial,  wodtke2011neighborhood}). And the recent stratification literature examines the effects of increased grandparent overlap on the grandchildren's status attainment (\citealt{lehti2018tying,song2019shared}). The logic of these focuses is straightforward - that a longer duration of overlap to a context shall amplify its impacts on people's social economic outcomes, be it a coresident father, teacher, neighborhood or grandparent.  
 
Although the idea is intuitive, overlap effects impose new conceptual challenges. From a sociological perspective, the overlap effect is a new category of estimands. An overlap time is special in that it bestows an effect only in a context. That is, for instance, the grandparent overlap effect is not derived from prolonging the grandparent's overlap devoid of the grandparent characteristics, but any effects of the grandparent characteristics which are amplified with a longer overlap should constitute the grandparent overlap effect, be it observed or unobserved characteristics. Since the characteristics of a person or a social context are inherently multi-dimensional, from a causal perspective, overlap effects should be multi-dimensional treatment effects. This conceptualization departs from the classical causal inference framework of no multiple version condition under the stable unit treatment value assumption (SUTVA) (\citealt{rubin1980discussion}), which postulates that the treatment effect should not vary depending on the different versions of the treatment received by an individual. In existing experimental and quasi-experimental designs of causal inference, this no multiple versions condition is usually translated into a uni-dimensional treatment, taking the form of a scalar intervention variable. While a uni-dimensional treatment framework may be sufficient in other settings, it may not grant sociologists the flexibility to capture the heterogeneous social processes of the overlap effects. 

In this paper, we propose a multi-dimensional framework of overlap effects. We formally define the overlap treatment as a vector which includes both the ``duration of overlap" and the ``content of overlap", the time-varying and time-invariant characteristics of the contexts. Correspondingly, we specify a  data generation model of multidimensional overlap effects in which one's outcome is determined by the accumulation of effects of time-varying and time-invariant characteristics of a context, either observable or unobservable, which can be confounders that affect the subsequent exposure to the context and the outcome. This model allows for, say, grandchild's outcome to be affected by the entire history of grandparent health and income and their unobserved culture and social capital over the course of overlap, and unobserved culture and social capital can affect the grandparent's health and mortality. This framework goes beyond the capacity of randomized-control trails(RCT) which is usually  infeasible to account for complicated treatment heterogeneity across multidimensional characteristics, and existing quasi-experimental designs, such as instrumental variable  estimator and fixed effects which cannot identify the multidimensional overlap effects given their effect heterogeneity across the unobservable. 

We propose a new estimator, the cumulative fixed effects (CFE), which recovers overlap effects from  three-wave individual panel data or from two-wave sibling data and grants applicants full control over the counterfactual stipulated contextual characteristics, i.e. the content of overlap, and hence greater specificity of the estimand. This regression-based approach includes four steps and applies to continuous outcomes. The first two steps eliminate the main effect of fixed unobserved confounders and their interactive effect with the duration of overlap in the model to identify the parameters of the observables. The third step recovers the part of the overlap effect drawn from the unobservables. The fourth step involves stipulating the values of the overlap content intervention, or the time-varying characteristics of the context during the extended overlap period. Our multi-step estimator has the heuristic and pedagogical advantage of mapping the statistical approach clearly to the estimands. In addition to the multi-step estimator, we also provide a generalized method of moment(GMM) estimation approach through which the estimation of overlap effect of interest can be readily achieved in a single step by specifying the right moment conditions (See Appendix A for details). Of course, this level of control over the estimand requires rich data. Fortunately, this kind of data is increasingly available from population registers. 
 
This operationalization of overlap effects has two advantages for sociological theory. First, our operationalization requires analysts to spell out the \textit{treatment heterogeneity} of their imagined counterfactual by specifying the hypothetical characteristics of the context in the additional period. This imbues the effects estimated by our approach with specificity and transparency regarding the implied counterfactual that might otherwise be swept under the rug by vague appeals to the causal effects of ``extending overlap". Second, it opens the door to principled reasoning about \textit{effect heterogeneity}, as overlap effects likely depend on the duration of overlap itself and the content. Consequently, we propose a suite of estimands designed to capture important aspects of this heterogeneity (i.e. length-specific overlap effects and conditional overlap effects) in addition to the population average overlap effect.

In the following sections, we discuss the grandparent overlap effect as a running example of overlap effects. Section 2 discusses the past framework of unidimensional overlap effect. Section 3 proposes the new notions of multidimensional overlap effects and several causal estimands of potential sociological interest. Section 4 illustrates our general estimands by specifying a theoretical model that links grandparent overlap to grandchild outcomes and thereby operationalizes our conceptualization of multidimensional overlap effects. Section 5 discusses the identification assumption. Section 6 introduces cumulative fixed effects estimation using a first-difference based approach and generalized methods of moments (GMM). Section 7 illustrates and evaluates our estimation approach by simulation. We discuss extensions and open issues in section 8. Section 9 concludes.     

\section{Unidimensional Overlap Effects}

We first show the definition, estimand and DGP of overlap effects under the unidimensional treatment framework.  In \cite{rubin1980discussion}'s classical framework of causal inference, the stable unit treatment value assumption (SUTVA) stipulates that there should be no multiple versions of the treatment  (\citealt{rubin1980discussion}). Under this condition, the treatment is usually standardized (i.e. the same drug or medical procedure is administrated to the entire intervention group), and the treatment effect shall not vary depending on the versions of the treatment an individual receives. In experimental and quasi-experimental designs, this no multiple versions condition is usually translated to a uni-dimensional treatment, namely, a scalar treatment variable. 

Under the uni-dimensional treatment framework, the overlap effects can be conceptualized as from the intervention on the ``duration" of overlap. Let $Y_{it}$ be the outcome of an individual  $i \in 1,...,N$ at time point $t\in 1,...,T$. The treatment is the duration of a contextual overlap, $A_{it}$, defined as the length of overlap exposure of $i$ to the social context up to time point $t$. 


We use potential outcomes notation to define the estimands (\citealt{rubin1974estimating}). Let $Y_{it}(a)$ be individual $i$'s outcome at time $t$ that would be observed if the individual had been exposed to overlap $A_{it}=a$. The individual-level effect of hypothetically increasing  $i$'s actual overlap by one year on her outcome at time $t$ is the individual overlap effect, $IOE_{it}^{uni}=\{Y_{it}(a_{it}+1)-Y_{it}(a_{it})|A_{it}<t\}$. Clearly, all effects of increasing overlap are only defined among individuals whose actual overlap duration is shorter than $t$, $A_{it}<t$, because, for example, an individual cannot have experienced $A_{it}+1=10$ years of overlap by age $t=9$. The population average overlap effect, $PAOE_{t}^{uni}$, defined as the average effect of increasing every individual's overlap by one year on their outcomes at time $t$, would be


\begin{align}
&PAOE_{t}^{uni}=E\big[Y_{it}(A_{it}+1)-Y_{it}(A_{it})| A_{it}<t\big].
\end{align}

To reflect how the duration of overlap $A_{it}$, contextual baseline characteristics $C_{i}$ and  contextual unobserved characteristics $U_{i}$ affects one's outcome $Y_{it}$, the data generating model can be:

\begin{align}
Y_{it}=\beta_0+\beta_{t} A_{it}+\alpha C_{i}+U_{i}+e_{it}
\label{eq:uni_dgp}
\end{align}

As usual, $\beta_0$ and $e_{it}$ represent the intercept and the idiosyncratic mean-zero error terms. 

It follows that

\begin{align}
&PAOE_{t}^{uni}=E\big[Y_{it}(A_{it}+1)-Y_{it}(A_{it})| A_{it}<t\big]=\beta_t.
\end{align}

$PAOE_{t}^{uni}$ is uni-dimensional in the sense that it only depends on one parameter on the overlap duration, $\beta_{t}$. We say that the $PAOE_t$ represents a class of estimands because it may vary according to the time point, $t$, at which one's outcome is assessed. 

This $PAOE_{t}^{uni}$ estimand is identifiable through eliminating the fixed effects term $U_i$ using sibling data and a conventional fixed effects estimator.\footnote{A sibling fixed effect strategy would require the data of a sibling $j$ for each individual $i$, for whom their duration of overlap, $A_{jt}$ and outcome, $Y_{jt}$, are measured, assuming that $i$ and $j$ share the same grandparental $C_i$ and $U_i$. For the sibling $j$, we have the data generating model of how overlap affects their outcome as:

\begin{align}
Y_{jt}=\beta_0+\beta A_{jt}+\alpha C_{i}+U_{i}+e_{it}.
\label{eq:uni_dgp2}
\end{align}
} In the literature of social mobility, \cite{lehti2018tying} adopted sibling fixed effects to exploit variations in the duration of grandparent overlap when both siblings reach age $t=20$ to identify the effect of grandparent overlap duration on grandchildren's educational attainment. With or without the fixed effect term $U_i$, such models of overlap duration are postulated in other sociological scenarios, in the study of duration of exposure to family complexity on children's outcomes (\citealt{gennetian2005one}), the neighborhood effect on teen's behaviors (\citealt{wodtke2011neighborhood,wodtke2013duration}) and teacher turnover effect on student's achievement (\citealt{ronfeldt2013teacher}).
 
While the unidimentional overlap effects $\beta_t$ are identifiable, they may not accurately capture the sociological reality of how contextual overlap shapes one's behavioral outcomes. Specifically, it rests on two simplifying assumptions which cannot easily be relaxed. First, the model assumes that the duration of overlap at $t$ is conditionally randomized at the older sibling's birth given the contextual baseline characteristics $C_i$ and $U_i$, which implies that contextual subsequent time-varying (i.e. sibling-varying) characteristics cannot affect the duration of overlap. This is problematic because some time-varying characteristics of the context may be confounders which affect both the outcome and the subsequent status of exposure. For instance, grandparent health is a time-varying confounder of grandparent overlap effect which affects both the grandparent's subsequent survival (and thus grandparent overlap) and the grandchildren's outcomes. The father's employment is a time-varying confounder which affects both the father's marital status and the children's outcome. A failure to control for these time-varying confounders would lead to a biased estimation of the overlap effect. But even when these time-varying confounders are observed, it is not obvious how they can be accounted for. In fact, these confounders are likely also mediators of the overlap effects, so controlling for them would risk controlling away part of the overlap effect. In short, the model of overlap duration cannot control for time-varying (sibling-varying) confounders because of the dual roles of the time-varying characteristics as both confounders and mediators,\footnote{In section 4, we will show that these time-varying characteristics can rather be conceptualized as part of the treatment under the multidimensional treatment framework.} resulting in a simplified depiction of social reality.

Second, the model implicitly assumes that the effect of overlap does not vary across contexts with different unobserved characteristics. If, for example the overlap effects are greater for a social context with higher  cultural and social capital, and if this heterogeneity is not explicitly modelled, then conventional fixed effects models at best recover variance-weighted average overlap effects (\citealt{wooldridge2005fixed, wooldridge2004fixed}). Unmodelled effect heterogeneity is of concern for two reasons. First,  variance-weighted average effects are not typically of interest to sociologists (\citealt{morgan2015counterfactuals}). Second, sociologists are specifically interested in how overlap effects vary across contextual characteristics, be it a school, a neighborhood, a grandparent or a father, as this effect heterogeneity contributes to the reproduction of social inequality (\citealt{mare2011multigenerational,pfeffer2014multigenerational, song2019shared}).

\section{Multidimensional Overlap Estimands}
\label{sec:est}

To better capture the social reality that overlap effects are cumulative effects of observed and unobserved contextual characteristics over the course of overlap, we propose a new framework of multidimensional overlap. We introduce the notion of multidimensional overlap and the corresponding estimands in this section. 

We define the multidimensional overlap, $\mathscr{A}_{it}$, as:

\begin{align}
\mathscr{A}_{it}=\{\bar C_{iA_{it}},U_i,A_{it}\}
\end{align}

$\mathscr{A}_{it}$ includes the duration of overlap, $A_{it}$, and the content of overlap, which includes two parts, first, $\bar C_{iA_{it}}=\{C_{i0},...,C_{it}\}$, the history of contextual observable time-varying characteristics from the start to the end of overlap $A_{it}$, and second, $U_i$, the contextual unobserved fixed characteristics.

Under this multidimensional treatment framework, the individual-level treatment effect is
\begin{align}
IOE_{it}=\{Y_{it}(\mathcal{a}_{it}^{+1})-Y_{it}(\mathcal{a}_{it})|A_{it}<t\} \\
where \; \mathcal{a}_{it}^{+1}=g\{\bar C_{ia_{it}+1}, U_{i}, a_{it}+1 \}\nonumber
\end{align}

$Y_{it}(\mathcal{a}_{it}^{+1})$ differs from $Y_{it}(a_{it}+1)$ in that it indicates the potential outcome of one's outcome given that the duration of overlap $a_{it}$ increases by a year and that the content of overlap, $\bar C_{it}$ and $U_i$, also extends for a year.

We propose two main classes of estimands and their extensions to answer several substantively interesting questions about overlap effects. The population average overlap effect, $PAOE_{t}$, is given by taking the expectation of $IOE_{it}$ over the population.

\begin{align}
&PAOE_{t}=E\big[Y_{it}(\mathscr{A}_{it}^{+1})-Y_{it}(\mathscr{A}_{it})| A_{it}<t\big].
\end{align}

The $PAOE_{t}$ will often be the primary object of interest, if for no other reason than that it provides a one-number summary of the potentially heterogeneous  overlap effects on individual's outcomes at a specific time, $t$. 

By the same token, however, the $PAOE_{t}$ likely averages over systematic and sociologically interesting effect heterogeneity. For example, the effect of overlap may change with the length of overlap because its content, $\bar C_{iA_{it}}$, changes. To account for possible non-linearities in overlap effects across the length of overlap, we define the length-specific average overlap effect, $LAOE_{t}(a)$, as the average effect of an additional year of overlap among individuals with a given length of overlap, $a$, on one's outcomes at time point $t$,

\begin{align}
&LAOE_{t}(a)=E\big[Y_{it}(\mathscr{A}_{it}^{+1})-Y_{it}(\mathscr{A}_{it})\big|A_{it}=a<t\big]
\end{align}

Both $PAOE_t$ and $LAOE_t(a)$ likely mask effect heterogeneity by the contextual baseline characteristics, which may be of most sociological interest. For instance, the extent of multigenerational transmission of (dis)advantages (\citealt{mare2011multigenerational,song2019shared}) can be estimated by the effect heterogeneity of $PAOE_t$ across grandparent's race, income or class. Such heterogeneity can be captured by conditioning either estimand by the relevant baseline characteristics. The conditional population-average overlap effect among the grandchildren with baseline characteristics $B_i=b$ is given by
\begin{align}
&CPAOE_{t}^b=E\big[Y_{it}(\mathscr{A}_{it}^{+1})-Y_{it}(\mathscr{A}_{it})\big|A_{it}<t, B_i=b\big],
\end{align}
and the conditional length-specific average overlap effect, $CLAOE_t^b(a)$, is given by 
\begin{align}
&CLAOE_t^b(a)=E\big[Y_{it}(\mathscr{A}_{it}^{+1})-Y_{it}(\mathscr{A}_{it})\big|A_{it}=a<t, B_i=b\big].
\end{align}
The definitions of the $IOE_{it}$, $PAOE_{t}$, and $LAOE_{t}(a)$, and their conditional extensions $CPAOE_{t}^b$, and $CLAOE_{t}^b(a)$, are generic, in the sense that they do not depend on any particular theory about the data generating process (DGP) of the overlap effects.


Table 1 summarises  the notation and definitions used throughout the paper, some of which will be introduced later.
\begin{center}
[Table 1 About Here]
\end{center}

\section{A Model for Grandparent Overlap Effects}
\label{sec:model}

Having defined overlap effects formally and generally as the effects of multidimensional exposure histories, we now apply this conceptual framework to the specific example of grandparent overlap effects. Since the identification of causal effects is always relative to a particular model of data generation, we first posit a flexible model that relates grandchildren’s test scores to the history of
grandparent characteristics and derive grandparent overlap effects. Subsequently, we will investigate the identification and estimation of grandparent overlap effects relative to this DGP. 

Overlap effects are multi-dimensional because they derive from cumulative effects of grandparent characteristics over the course of overlap. Hence, we need a cumulative model to reflect their multidimensional nature. We define grandparent overlap effects as the cumulative effects of grandchildren's exposure to grandparent's observed and unobserved characteristics across the duration of overlap. 

\begin{figure}[!htbp]
		\centering
			\begin{minipage}{1\linewidth}
		\includegraphics[width=\linewidth]{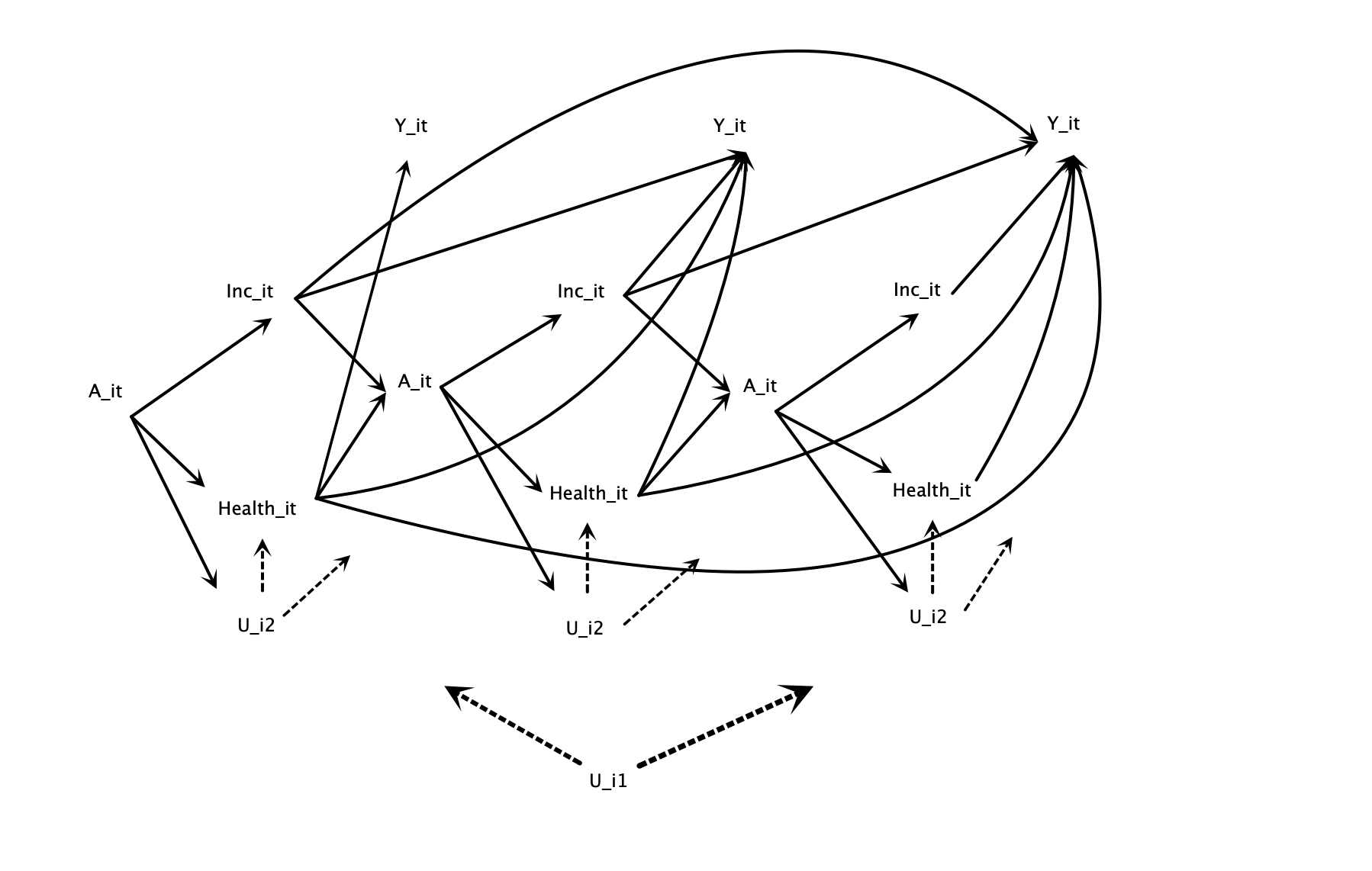}
			\caption{Causal diagram of overlap duration, overlap context and outcome}
		
\footnotesize		
\emph{Notes:} $A_{it}$ represents the duration of overlap at $t$. $Y_{it}$ is the grandchild's outcome. Without loss of generality, $C_{it}=(Inc_{it}, Health_{it})$, are examples of grandparent time-varying characteristics, i.e. the overlap context. To simplify the visualization, dashed lines are used to represent the effect of $U_{i1}$ and $U_{i2}$ on all the observables, and only three waves are shown. 
	\label{fig:DAG_mul}
	\end{minipage}
\end{figure}

We first present a causal diagram to describe our theory of how grandparents characteristics and overlap affect grandchildren's test scores (see the DAG in Figure \ref{fig:DAG_mul}). The DAG describes the causal relationships of all variables, including the mediation relationships among $A_{it}$, $C_{it}$ (take for instance grandparent income, $Inc_{it}$, and grandparent health, $Health_{it}$) and $U_{i2}$ which are not part of the model of the outcome.

A linear model of the outcome may take this following form:

\begin{align}
	Y_{it}&=\beta_{0}+\sum_{t'=1}^{A_{it} }(C_{it'}\beta_{t'}+U_{i2})+U_{i1}+\epsilon_{it}\\
	&=\beta_{0}+\sum_{t'=1}^{A_{it} }(C_{it'}\beta_{t'})+U_{i2}A_{it}+U_{i1}+\epsilon_{it}
	\label{eq:dgp}
\end{align}

In this model, grandchild test scores at age $t$, $Y_{it}$, are affected by grandparent's time-varying and time-invariant characteristics, $C_{it'}$, $U_{i1}$ and $U_{i2}$, across the entire course of overlap, from $t'=1$ to $A_{it}$. As usual, $\beta_0$ and $\epsilon_{it}$, represent the intercept and idiosyncratic, period-specific, mean-zero error terms. As a key innovation, we recognize that the grandparent fixed characteristics (e.g. education, values, genes, and class background) may exert both a constant (``fixed") effect on grandchild's test scores at a given child's age, captured by the term $U_{i1}$, and an effect that accumulates with overlap, captured by the term $U_{i2}A_{it}$ (``cumulative fixed effect''). For example, the grandparent's class position may afford the grandchild a fixed bonus of motivation regardless of the grandparent's survival status, $U_{i1}$, and each additional year of grandchild's interaction with the exemplars of the family's privilege may confer additional advantage via building grandchild's sense of entitlement, $U_{i2}$. The model remains agnostic whether any given fixed characteristic exerts both a fixed and a cumulative effect, only a fixed effect, or only a cumulative effect, i.e. whether it is an element of $U_{i1}$, $U_{i2}$, or both. The heart of the model resides in the middle three terms- that is, the total grandparent overlap is generated by the two terms containing the duration of overlap, $A_{it}$, namely $\sum_{t'=1}^{A_{it}}C_{it'}\beta_{t'}$ and $\sum_{t'=1}^{A_{it}}U_{i2}$, which we call the observable and unobservable components of the grandparent overlap effect, respectively. Since $U_{i2}$ is time-invariant,  $\sum_{t'=1}^{A_{it}}U_{i2}=U_{i2}A_{it}$. \footnote{In contrast to $U_{i2}$, $U_{i1}$ is a grandparent effect but not part of the grandparent overlap effect, since it captures the grandparental influence that would exist even if the grandparent had died before grandchild's birth, $A_{it}=0$. $U_{i1}$ may include, for instance, the grandparent wealth.} 
The term $\sum_{t'=1}^{A_{it}}C_{it'}\beta_{t'}$ captures the effect of the history of grandparent's observed time-varying characteristics, $\bar C_{it}=\{C_{i1},...,C_{iA_{it}}\}$, from grandchild's birth at $t'=1$ across the length of overlap, $A_{it}$, where $t'$ is the index of summation.  Time-varying effects, $\beta_{t'}$, acknowledge, that a given grandparental characteristic may have different effects depending on the age at which the grandchild experienced the characteristic. \footnote{For example, if $C_{it'}$ is a time-varying indicator of grandparent's ill health at grandchild's age $t'$, and $\beta_2\neq\beta_{15}$, then grandparent's illness at grandchild's age $t'=2$ has a different effect on grandchild's outcome at age $t$ than does grandparent's illness at grandchild's age $t'=15$.}

Compared to the unidimensional model in equation \ref{eq:uni_dgp}, the multidimensional overlap model engages the entire social processes after the grandchild's birth throughout the entire course of overlap. There are three features of the multidimensional model which highlight its flexibility but also underlie its identification and estimation challenges. First, the causal paths of $A_{it}$ on $Y_{it}$ at each $t$ are entirely mediated by the grandparent observed characteristics $C_{it}$ (i.e. $C_{1it}$ and $C_{2it}$) and unobserved characteristics $U_{i2}$. This represents a special case of causal mediation effect in which treatment effect is entirely mediated by the mediators without direct effect of the treatment. This causal structure is consistent with \cite{vanderweele2013causal}'s DAG of multiple versions of the treatment where the effect of the treatment (e.g. the treatment of receiving a surgery) is entirely absorbed by the versions of the treatment (e.g. the specific surgeons who conduct the surgery). The notion of a ``version" in \cite{vanderweele2013causal} corresponds to a unique combination of the grandparent characteristics in our multi-dimensional overlap effect case. Correspondingly,  the parametric DGP does not include a ``main effect" for overlap, $A_{it}$. This visible departure from past sociological research on overlap effects is intentional. Overlap $A_{it}$ (i.e. the duration) cannot exert an effect on grandchild test scores net of these characteristics (i.e. the content), and hence does not merit a main-effects term. 

Second, unlike the unidimensional model where only grandparent baseline characteristics are allowed, the multidimensional model permits grandparent time-varying characteristics, $C_{it'}$. At each period $t'$, $C_{it'}$ is a vector of grandparent's characteristics that affect grandchild's test scores at $t$, including, for example, grandparent's health, income, labor force participation, marital status, and coresidence with the grandchild (only two $C_{it'}$ are shown in the DAG for simplicity). \footnote{Readers may notice that the model does not explicitly include time-invariant contextual characteristics that are observable. When the observed time-invariant characteristics are believed to have a fixed effect on the outcome, they are a part of $U_i=(U_{i1}, U_{i2})$ . When they are believed to have a time-varying effect on the outcome, for instance, the fixed grandparent education may contribute more as the grandchild gets older, then the time-invariant observables  can be regarded as a special case of time-varying observables $C_{it'}$ with zero variance across $t'$.} The model is cumulative in that all past $C_{it'}$ are allowed to have an effect on any subsequent children's outcome $Y_{it}$. Besides, grandparent's past characteristics, $C_{it'-s}$, $s>0$, are permitted to affect grandparent's present characteristics $C_{it'}$ via $A_{it'}$ (the survival status). 

Third, $C_{it'}$ and $U_{i2}$ play the dual roles of confounders (which affect both the subsequent outcome and treatment) and the versions of the treatment, which impose unseen challenges of identification and estimation on existing approaches. As confounders, $C_{it}$ and $U_{2}$ need to be adjusted in order to identify the treatment effect of $A_{it}$. But since they also constitute the treatment effect, and the grandparents need to survive in order to have another wave of measures of $C_{it}$,  $C_{it}$ cannot be controlled away if sociologists are interested in recovering the total overlap effects. For this reason, conventional fixed effects models and fixed effects individual slope models (\citealt{ruttenauer2020fixed}) cannot identify or estimate the theoretical model in figure \ref{fig:DAG_mul}. 

The individual-level causal effect of one additional year of grandparent overlap on grandchild $i$ in our model is given by 

\begin{align}
IOE_{it}=\{Y_{it}(\mathscr{A}_{it}^{+1})-Y_{it}(\mathscr{A}_{it})|A_{it}<t\}=\{C_{ia_{it}+1}\beta_{a_{it}+1}+U_{i2}|A_{it}<t\}.
\label{eq:indivi_effect}
\end{align}

The $LAOE_t(a)$ gives the average grandparent overlap effect among children who experience a given length of overlap, $A_{it}=a$, 
\begin{align}
LAOE_t(a)&=E\big[Y_{it}(\mathscr{A}_{it}^{+1})-Y_{it}(\mathscr{A}_{it})| A_{it}=a<t\big]\nonumber\\
&=E\big[C_{iA_{it}+1}\beta_{A_{it}+1}+U_{i2}| A_{it}=a<t\big].
\label{eq: LAOE}
\end{align}
The $PAOE_{t}$ averages across all $IOE_{it}$'s in the entire population,\footnote{The $PAOE_{t}$ is also the average of the $LAOE_t(a)$ s, $E_{A_{it}}\Big[E\big[Y_{it}(a+1)-Y_{it}(a)\,|\,A_{it}=a<t\big]\Big]$.} 
\begin{align}
PAOE_{t}&=E\big[Y_{it}(\mathscr{A}_{it}^{+1})-Y_{it}(\mathscr{A}_{it})|A_{it}<t\big]\nonumber\\
&=E\big[C_{iA_{it}+1}\beta_{A_{it}+1}+ U_{i2}| A_{it}<t\big]
\label{eq:PAOE}
\end{align}

We add two remarks. First, the content of overlap, the grandparent characteristics during the hypothetically granted additional year of life, $C_{iA_{it}+1}$ and $U_{i2}$, can be thought of as the ``versions" or mediators in the analysis of controlled direct effect (\citealt{vanderweele2015explanation}).  And since the overlap effect is entirely mediated by the content of overlap, we could sum all the controlled direct effects to recover the total effect of overlap. In other words, when defining the effect of hypothetically prolonging a grandparent's life, the analyst owes the reader a statement as to what values each element of the vector of $C_{iA_{it+1}}$ is supposed to have. 


For some classes of grandparent deaths (e.g., freak traffic accidents), it is easy to imagine a counterfactual scenario in which the grandparent would naturally live another year in good health. In other cases (e.g., death from old age), however, the natural counterfactual would be living at death's door, perhaps in a coma. Instead of hiding the implicit counterfactual, our approach thus empowers--indeed, requires--analysts to specify the desired grandparental characteristics as additional hypothetical interventions, which conventional estimation approaches typically do not allow. \footnote{The requirement of clearly stating the counterfactual is not unique to conceptualizing grandparent overlap effects. For example, in studies of the effect of parental overlap on child outcomes, analysts should similarly specify whether they imagine the counterfactual parental relationship to divorce to be ``happy marriage" or ``marriage at the verge of divorce."} That being said, analysts should pause to consider which values of $C_{iA_{it}+1}$ are both interesting and plausible. To this end, they might want to predict (or even structurally model) these characteristics from the history of grandparents' prior characteristics, as discussed below.

Second, honoring the growing sociological interest in effect heterogeneity (\citealt{brand2013causal, elwert2010effect,elwert2014endogenous,xie2013population}), our model allows individual-level grandparent overlap effects to vary for several reasons: The $IOE_{it}$s may vary across grandchildren, $i$, at a fixed age, $t$, because (1) grandparents may have different characteristics, $C_{ia_{it}+1}$, in the hypothetically granted additional year of life, (2) the effect of these characteristics may vary with the duration of overlap since $\beta_{a_{it}+1}$ varies with $a_{it}$, and (3) grandparents may vary with respect to their fixed characteristics, $U_{i2}$. The $LAOE_t(a)$ and $PAOE_{t}$ estimands simply average across these heterogenous individual-level effects in different ways but retain variation due to grandparents' observed and unobserved characteristics.

Like all models, the DGP that we have analyzed in this section contains simplifications.  First, as discussed below in section \ref{sec:assum}, the model rules out unobserved time-varying confounders---a property it shares with conventional panel fixed effects models. Second, it asserts that the cumulative effects of grandparent fixed unobserved characteristics are linear in overlap, $U_{i2}A_{it}$, although this assumption could easily be relaxed with respect to the observed characteristics in $U_{i2}$. Third, the model imposes a restriction on how past grandparental characteristics may affect test scores in the present via $\beta_{t'}$, in that the effect of grandparent inputs at a given age, $t'$, is not allowed to decay over time, $t$, although we consider alternative specifications in \autoref{sec:gap_varying}. 

\section{Identification}
\label{sec:assum}

The parameters of the overlap model in \autoref{eq:dgp} are identified under the assumption of strict exogeneity. Remarkably, this is the same assumption that also underlies the identification of conventional panel fixed effects models that do not contain cumulative fixed effects (\citealt{chamberlain1982multivariate}). 
Consider a balanced panel of $i \in 1,...,N$ grandchildren observed for $t' \in 1,...,T$ periods. The strict exogeneity assumption,
\begin{align}
&E(\epsilon_{it'}|\bar C_{iA_{iT}}, A_{iT}, U_i)=0,&
\label{eq:assumption}
\end{align}

states that, for each grandchild $i$ and each period $t'$, the idiosyncratic period-specific error term, $\epsilon_{it'}$, is conditionally mean independent of (1) the past and future history of contextual observed time-varying characteristics across the entire period of overlap, $\bar C_{iA_{iT}}$, (2) overlap itself, $A_{iT}$, (i.e. grandparent vital status in each period), and (3) the fixed characteristics, $U_i=(U_{i1}, U_{i2})$. 

Using the potential outcome framework, if we use the notation of $A\independent B|C$ to note that A is independent of B given C, the same condition of equation \ref{eq:assumption} would be translated to:

\begin{align}
& Y_{it}(\mathcal{a}_{it})\independent \mathscr{A}_{it}|\bar C_{iA_{iT}}, U_i
\end{align}

Substantively, this assumption says that, 1) past outcomes may not directly affect the current exposures, and 2) there are no unobserved time-varying confounders after conditional on the contextual observed covariate history, $\bar C_{iA_{it}}$, and the fixed effects (\citealt{sobel2012does, imai2019should}). To what extent the assumption is plausible depends on the empirical application and should be judged on a case-by-case basis, accounting for the specific outcome of interest and the available overlap content treatments which are observed confounders and the causal relationships among them. 

In illustrating how this identification assumption unfolds for the example of grandparent overlap effect on grandchildren's cognitive test scores, the first condition can be translated into that grandchildren's past academic test scores do not cause grandparent's survival, $A_{it}$, or the content of grandparent overlap, $C_{it'}$, such as income, or health. This condition may be plausible given that intergenerational transfers of interpersonal and financial resources and transmission of status goes downstream from the grandparent generation to the grandchild's generation, especially when the grandchild is younger. But analysts shall discuss the implications if theories or findings show that grandchildren's cognitive outcomes can form a feedback loop to affect the grandparents well-being or survival. The second claim regarding the absence of time-varying unobservable confounders will have to be justified by arguing that the covariates, $\bar C_{it'}$, observed in a particular empirical application are sufficiently detailed to control for time-varying confounding. This condition should be evaluated given the data at hand. 



Notably, analysts should and could control for time-varying variables which are confounders of the outcome and the treatment variable even though they are not part of the content of overlap. In the case of grandparent overlap effect, it may include parent or grandchild characteristics which are not themselves part of the grandparent effect, for instance, the parent income and marital status. But analysts should not control for the mediators of content of overlap and may (i.e. the parent mediators on the causal pathway between past grandparent characteristics and future grandchild test scores) and otherwise, the analysis will incur overcontrol bias (\citealt{elwert2013graphical, elwert2014endogenous}). 


\section{The Cumulative Fixed Effects Estimator}
\label{sec:hetero}
Under the strict exogeneity assumption, the $PAOE_{t}$ and $LAOE_t(a)$ generated by the model in \autoref{eq:dgp} are identifiable over some range of overlap durations from individual-level panel data with at least three waves of outcomes and detailed histories of grandparent characteristics.\footnote{Alternatively, one could estimate the overlap effects from panel data on sibling pairs with two waves of outcomes. The extension is straight-forward. We focus on individual-level panels henceforth.} Estimation involves four steps. The first two steps exploit within-individual variation in the outcome to remove the main effect of the unobservables, $U_{i1}$, and their interaction with grandparent overlap, $U_{i2}A_{it}$, respectively. These steps identify the parameters, $\beta_{t'}$, on the observed time-varying grandparent characteristics, $C_{it'}$. The third step recovers the contribution of the fixed unobservables, $U_{i2}$, to the grandparent overlap effect. In the fourth step, the analyst chooses the time-varying characteristics, $C_{iA_{it}+1}$, that the grandparent should posses during their additional year of life. With these inputs, one can then compute the $LAOE_t(a)$ and $PAOE_{t}$ of an additional year of overlap with the chosen characteristics for certain durations of overlap. We also provide a overarching generalized method of moment (GMM) estimation procedure to recover the parameters and the overlap effects ($PAOE_{t}$ and $LAOE_t(a)$) directly, which allows analysts to use any modern GMM package(e.g. in Stata or R) with minimum programming (See Appendix A for details).

\subsection{Data Structure}
Consider individual-level panel data that follow one cohort of grandchildren and their grandparent from the time of grandchildren's birth, $t'=1$, until the outcome, $Y_{it}$, has been measured three times. Let the grandchildren's age at the three outcome assessments (``test dates") be $q_1<q_2<q_3$, where the third test date can effectively be thought of as the end of follow up, $q_3=T$. The test dates are shared across grandchildren in the cohort but do not need to be equally spaced.  For reasons that will become apparent shortly, we restrict the analysis to grandchildren whose grandparent does not die until after the second test date, $d_i>q_2$. Therefore, $A_{iq_1}=q_1$, $A_{iq_2}=q_2$, and $q_2<A_{iq_3}\leq q_3$. Examples of suitable data sets might include research panels or administrative data linking families across generations in areas that maintain regular educational testing programs, like those increasingly available in some states of the United States, Europe, or Asia.

\subsection{Steps 1 and 2: Estimation of $\beta_{t'}$}

We start by transforming the data in order to eliminate the dependence of the outcome on the unobserved confounders. Since our model permits that the unobservables exert not only a conventional fixed effect, $U_{i1}$, on the outcome, but also interact with the duration of overlap in a cumulative fixed effect,  $U_{i2}A_{it}$, the required transformations are somewhat more involved than in conventional panel fixed effects estimation. 

The first step takes the conventional first difference of \autoref{eq:dgp} between consecutive outcome waves twice, in order to eliminate the main effect of the unobservables, $U_{i1}$, yielding
\begin{align}
&\bigtriangleup_{q_2q_1}Y_{i}=\sum_{t'=q_1+1}^{A_{iq_2}}C_{it'}\beta_{t'}+U_{i2} \bigtriangleup_{q_2q_1}A_{i}+\bigtriangleup_{q_2q_1}\epsilon_{i},
\label{eq:first_diff}
\end{align}
and
\begin{align}
&\bigtriangleup_{q_3q_2}Y_{i}=\sum_{t'=q_2+1}^{A_{iq_3}}C_{it'}\beta_{t'}+U_{i2} \bigtriangleup_{q_3q_2}A_{i}+\bigtriangleup_{q_3q_2}\epsilon_{i},
\label{eq:first_diff2}
\end{align}
where $\Delta$ indicates the difference between any variable $V_i$ at two different time points $a$ and $b$, and that $\Delta_{ab}V_i=V_{ia}-V_{ib}$. These first-differenced equations, cannot yet be consistently estimated by regression, because the terms involving $U_{i2}$ are unobserved confounders.  

The second step therefore removes the dependence on $U_{i2}$ by first rescaling and then differencing \autoref{eq:first_diff} and \autoref{eq:first_diff2}. Specifically, we first divide \autoref{eq:first_diff} by $\bigtriangleup_{q_2q_1}A_{i}$,

\begin{align}
&\frac{\bigtriangleup_{q_2q_1}Y_{i}}{\bigtriangleup_{q_2q_1}A_{i}}=\frac{\sum_{t'=q_1+1}^{A_{iq_2}}C_{it'}\beta_{t'}}{\bigtriangleup_{q_2q_1}A_{i}}+U_{i2}+\frac{\bigtriangleup_{q_2q_1}\epsilon_{i}}{\bigtriangleup_{q_2q_1}A_{i}},
\label{eq:firstdiffdivide A}
\end{align}
then divide \autoref{eq:first_diff2} by $\bigtriangleup_{q_3q_2}A_{i}$,
\begin{align}
&\frac{\bigtriangleup_{q_3q_2}Y_{i}}{\bigtriangleup_{q_3q_2}A_{i}}=\frac{\sum_{t'=q_2+1}^{A_{iq_3}}C_{it'}\beta_{t'}}{\bigtriangleup_{q_3q_2}A_{i}}+U_{i2}+\frac{\bigtriangleup_{q_3q_2}\epsilon_{i}}{\bigtriangleup_{q_3q_2}A_{i}},
\label{eq:firstdiffdivide A2}
\end{align}
and finally subtract \autoref{eq:firstdiffdivide A2} from \autoref{eq:firstdiffdivide A} , 
\begin{align}
&\frac{\bigtriangleup_{q_2q_1}Y_{i}}{\bigtriangleup_{q_2q_1}A_{i}}-\frac{\bigtriangleup_{q_3q_2}Y_{i}}{\bigtriangleup_{q_3q_2}A_{i}}=\frac{\sum_{t'=q_1+1}^{A_{iq_2}}C_{it'}\beta_{t'}}{\bigtriangleup_{q_2q_1}A_{i}}-\frac{\sum_{t'=q_2+1}^{A_{iq_3}}C_{it'}\beta_{t'}}{\bigtriangleup_{q_3q_2}A_{i}}+\frac{\bigtriangleup_{q_2q_1}\epsilon_{i}}{\bigtriangleup_{q_2q_1}A_{i}}-\frac{\bigtriangleup_{q_3q_2}\epsilon_{i}}{\bigtriangleup_{q_3q_2}A_{i}}.
\label{eq:second_diff}
\end{align}

\autoref{eq:second_diff} no longer depends on $U_{i1}$ or $U_{i2}$. Furthermore, the two terms involving the differenced error terms, $\epsilon_{it}$, are mean independent of the terms involving covariates, $C_{it'}$, under the assumption of strict exogeneity (\autoref{eq:assumption}). Therefore, the $\beta_{t'}$ parameters in \autoref{eq:second_diff} can be consistently estimated by OLS regression. 

We add two remarks. First, it is obvious that this estimator requires that grandparents survive beyond the second test date. Otherwise, \autoref{eq:firstdiffdivide A2} would involve dividing by $\bigtriangleup_{q_3q_2}A_{i}=0$. Second, the estimator only uses information between the first and third test date, $q_1$ and $q_3$. Therefore, it can only recover the parameters $\beta_{t'}$ for observed grandparent characteristics between these dates, $t'\in[q_1+1,q_3)$. Clearly, this is not much of a limitation if the tests are widely spaced apart. 

Notably, the two steps shown above are not the only way to the estimation of $\beta_t'$. In fact, fixed effects individual slope models (\citealt{ruttenauer2020fixed}) can also be used to achieve this goal. Estimation of the $PAOE_{t}$ and the $LAOE_t(a)$, however, requires not only the coefficients for the grandparent characteristics, $\beta_{t'}$, but also knowledge of the observable and unobservable components of the grandparent overlap effect, defined in \autoref{sec:model}. We turn to their estimation next.

\subsection{Step 3: Estimation of the Unobservable Component}

Estimating the contribution of the unobservable components--$E(U_{i2}|A_{it}<t)$ for the $PAOE_t$ and $E(U_{i2}|A_{it}=a<t)$ for the $LOAE_t(a)$--is difficult.  Recognizing that these components are part of the residuals, however, we can recover them between the second and third test date,  $E(U_{i2}|q_2<A_{it}<q3)$ and $E(U_{i2}|A_{it}=a, q_2<A_{it}<q_3)$. This implies that we can identify the $PAOE_t$ and $LAOE_t(a)$ only for durations of overlap corresponding to deaths between the second and third test date.

Subtracting the observables from \autoref{eq:first_diff2} yields
\begin{align}
&\bigtriangleup_{q_3q_2}Y_{i}-\sum_{t'=q_2+1}^{A_{iq_3}}C_{it'}\hat\beta_{t'}=U_{i2} \bigtriangleup_{q_3q_2}A_{i}+\bigtriangleup_{q_3q_2}\epsilon_{i}.
\label{eq:residuals}
\end{align}
Taking the conditional expectation gives

\begin{align}
&E\big(\bigtriangleup_{q_3q_2}Y_{i}-\sum_{t'=q_2+1}^{A_{iq_3}}C_{it'}\hat\beta_{t'}\big|A_{iq_3}, \bar C_{iA_{iq_3}}\big) \nonumber\\
&=E\big(U_{i2} \bigtriangleup_{q_3q_2}A_{i}\big|A_{iq_3}, \bar C_{iA_{iq_3}}\big)+E\big(\bigtriangleup_{q_3q_2}\epsilon_{i}\big|A_{iq_3}, \bar C_{iA_{iq_3}} \big).
\label{eq:CEresiduals}
\end{align}
Notice that $E(\bigtriangleup_{q_3q_2}\epsilon_{i}| A_{iq_3}, \bar C_{iA_{iq_3}})$=0 under the strict exogeneity assumption. Dividing equation \autoref{eq:CEresiduals} by $\bigtriangleup_{q_3q_2}A_{i}$ and rearranging terms expresses the conditional residual in terms of the observables,
\begin{align}
E(U_{i2}|A_{iq_3}, \bar C_{iA_{iq_3}})= E\frac{\big(\bigtriangleup_{q_3q_2}Y_{i}-\sum_{t'=q_2+1}^{A_{iq_3}}C_{it'}\hat \beta_{t'}\big|A_{iq_3}, \bar C_{iA_{iq_3}}\big)}{A_{iq_3}-A_{iq_2}}.
\label{eq:CEres_final}
\end{align}

Marginalizing over the conditioning arguments then gives the desired expressions. 

For example, consider a model where $C_{it'}$ includes only two grandparent characteristics, income, $I_{it'}$, and health, $H_{it'}$. Recalling that the $PAOE_{t}$ is defined only for $A_{it}<t$, we would use \autoref{eq:CEres_final} to estimate the unobservable term $E\big[U_{i2}|q_2<A_{iq_{3}}<q_3\big]$ in the $PAOE_t$  as
\begin{align}
\frac{1}{m}\sum_{i=1}^m\big(\frac{Y_{i,q_3}-Y_{i,q_2}-\sum_{t'=q_2+1}^{t'=A_{iq_3}}I_{it'}\hat \beta_{t'}-\sum_{t'=q_2+1}^{t'=A_{iq_3}} H_{it'}\hat\beta_{t'}}{A_{i,q_3}-q_2} \big),
\end{align}
where $m$ is the number of grandchildren in the sample whose grandparent die after $q_2$ and before $q_3$. The analogous estimate for the $E\big[U_{i2}|A_{i,q_3}=a, q_2<A_{iq_{3}}<q_3\big]$ term in the $LAOE_t(a)$ would be
\begin{align}
\frac{1}{m_a}\sum^{i \in A_{iq3}=a}\big(\frac{Y_{i,q_3}-Y_{i,q_2}-\sum_{t'=q_2+1}^{t'=a}I_{it'}\hat \beta_{t'}-\sum_{t'=q_2+1}^{t'=a} H_{it'}\hat\beta_{t'}}{a-q_2} \big),
\end{align}
 where $m_a$ is the number of observations of the subgroup with the length of overlap at the third test equal to $a$, i.e. $A_{i,q_3}=a$.

\subsection{Step 4: Choosing Grandparent Characteristics During the Extended Overlap Period}

As a last step, we need to choose values for the observable components: $E\big[C_{iA_{iq_{3}}+1}|A_{iq_{3}}=a, q_2<A_{iq_{3}}<q_3]$ for the $PAOE_t$, and $E\big[C_{iA_{iq_{3}}+1}| q_2<A_{iq_{3}}<q_3]$ for the $LOAE_t(a)$, respectively. Because grandparents' characteristics are only measured until they die, the grandparent characteristics in the extended overlap period, $C_{iA_{iq_{3}}+1}$, are in nature hypothetical and unobserved. We regard the mean of the grandparent characteristics $E(C_{iA_{iq_{3}}+1}| A_{iq_{3}}=a, q_2<A_{iq_{3}}<q_3)$ as intervention variables, whose values have to be chosen by the analyst in order to explicate the precise counterfactual to having a dead grandparent. 
 
Analysts have wide latitude in choosing theoretically interesting and substantively plausible values for $E\big[C_{iA_{iq_{3}}+1}|A_{iq_{3}}=a, q_2<A_{iq_{3}}<q_3]$ and $E\big[C_{iA_{iq_{3}}+1}| q_2<A_{iq_{3}}<q_3]$. Here, we list three different approaches for illustration. First, sociologists could ask what the grandparent overlap effect would be if the deceased grandparent's characteristics were to remain unchanged from the last period prior to grandparent's actual time of death, $C_{iA_{iq_{3}}+1}=C_{iA_{iq_{3}}}$. Indeed, when the time series of the grandparents' time-varying characteristic are mean stationary, the last observation would be an unbiased prediction of the future realization (\citealt{hyndman2018forecasting}), i.e. $\hat E\big[C_{iA_{iq_{3}}+1}| q_2<A_{iq_{3}}<q_3\big]=E\big[C_{iA_{iq_{3}}}|q_2<A_{iq_{3}}<q_3\big]$ and $ \hat E\big[C_{iA_{iq_{3}}+1}\big| A_{iq_{3}}=a, q_2<A_{iq_{3}}<q_3\big]=E\big[C_{iA_{iq_{3}}}\big| A_{iq_{3}}=a, q_2<A_{iq_{3}}<q_3\big]$.

Second, when the time series of grandparent time-varying characteristics are not stationary, analysts might assume that the average characteristics in the extended overlap period of the grandparents who die at $a$ would equal the average characteristics of the grandparents who die in the subsequent period, $a+1$. For example, $\hat E\big[C_{iA_{iq_{3}}+1}\big| A_{iq_{3}}=a, q_2<A_{iq_{3}}<q_3 \big]=\frac{1}{m_{a+1}}\sum_{i\in A_{iq3}=a+1} C_{iA_{iq3}+1}$, where $m_{a+1}$ is the number of grandparents whose length of overlap is $A_{it}=a+1$.
 
Third, to better reflect each grandparent's individual life-course trajectory, sociologist could use the history of grandparent's time-varying characteristics, $\bar C_{iA_{iq_{3}}}$, to predict $C_{iA_{iq_{3}}+1}$, for example using auto-regressive models or machine learning approaches. 
 
Various methods of forecasting the grandparent's future characteristics (if they had stayed alive) are available. Different assumptions about the time-series of grandparent characteristics will lead to different estimates for the various grandparent overlap effects. It is an advantage that our method remains agnostic about the process that would generate future grandparent characteristics. Whichever the choice of future grandparent characteristics that the analyst wishes to defend as realistic or informative, it can be plugged into the estimator. What is more, analysts could even entirely eschew the prediction of future grandparent characteristics and freely choose desired reference values to answer hypothetical questions, e.g., about the effect of another year with a healthy grandparent or a rich grandparent. The choice of the research question--via the choice of the counterfactual characteristics of the grandparent--is not mandated by nature, but is the choice of the researcher (\citealt{lundberg2021your}). 
 
 Finally, in order to estimate the $PAOE_t$ and $LAOE_t(a)$, we substitute the estimated parameters $\beta_{t'}$, the estimated unobservable components $E[U_{i2}|q_2<A_{iq_{3}}<q_3]$, $E[U_{i2}|A_{iq_{3}}=a, q_2<A_{iq_{3}}<q_3]$, and the chosen values for the observable components $E\big[C_{iA_{iq_{3}}+1}|A_{iq_{3}}=a, q_2<A_{iq_{3}}<q_3]$ and $E\big[C_{iA_{iq_{3}}+1}|q_2<A_{iq_{3}}<q_3]$ into  \autoref{eq:PAOE} and \autoref{eq: LAOE}. Estimates of $PAOE_t$ and $LAOE_t(a)$
will be unbiased given the chosen values of $C_{A_{iq_{3}}+1 }$, as long as the estimates of $\beta_t$ and $E[U_{i2}|q_2<A_{iq_{3}}<q_3]$ and $E[U_{i2}|A_{iq_{3}}=a, q_2<A_{iq_{3}}<q_3]$ are unbiased. Standard errors should be computed using bootstrap methods.

In sum, one can estimate $LAOE_t(a)$ and $PAOE_t$ following the four steps introduced above. The above estimation strategy is easily extended to the estimation of conditional grandparent overlaps effects, $CPAOE_t^b$ and $CLAOE_t^b(a)$, which capture effect heterogeneity across groups defined by the family's fixed or baseline characteristics, $B_i$. The easiest way is to apply the four steps on stratified subsamples defined by $B_i$. We illustrate the results by simulations.

\section{Simulation}
\label{sec:simu}

To demonstrate our approach, we specify a data generating model for the effect of grandparent overlap on grandchild test scores with known parameters and evaluate the finite-sample performance of our estimator by simulation. This section outlines the simulation procedure and presents key results. Details are shown in Appendix B. 

\subsection{Data Generation}

We specify a data-generating process that evolves in annual increments from grandchildren's birth to age 20. For simplicity, each grandparent has exactly one grandchild. We first generate exogenous inputs for each grandparent-grandchild pair as i.i.d. draws from various distributions: The grandparent-level fixed unobservables that confound grandparent overlap and grandchild test scores are drawn from $U_i \sim N(10,100)$ and $U_{i1}=U_{i2}$. The idiosyncratic age-specific errors in grandchildren's test scores are drawn from $\epsilon_{it}\sim N(0,10)$. Grandparents' age at grandchild's birth is uniformly distributed between ages 50 and 60 as a function of $U_i$, since mothers' (and hence grandmothers') age of childbearing correlates with their latent socio-economic status (\citealt{fomby2014age}).

Next, we posit two time-varying grandparental characteristics, $C_{it'}=\{I_{it'},H_{it'}\}$, income and health, that co-evolve as a function of (i) each other's values at $t'=1$, $C_{i1}$, which are themselves function of grandparent age; (ii) each other's most recent values, $C_{it'-1}$; (iii) the fixed effects, $U_i$; and (iv) their own time trends. The parameters of the process are tuned such that grandparents' health decreases over time and grandparents' income increases over time within individuals, with increasing variance across individuals, as is common in western countries (\citealt{deaton1994intertemporal}). The resulting time-series are non-stationary, i.e., have changing means and variances over time. 

Grandparent overlap, $A_{it'}$, is endogenously determined by all of the above inputs. We generate grandparent's survival so that it negatively depends on grandparents' age and positively depends on their income and good health. 

Finally, from these variables, we generate grandchild test scores, $Y_{it}$ at three test dates, $q_1=6$, $q_2=10$ and $q_3=20$, according to the reduced-form DGP of \autoref{eq:dgp}. We posit that grandparents' observed characteristics measured at times $t'$ affect grandchildren's test scores with parameters $\beta_{t'I}=\beta_{t'H}=5+0.5*(t'-1)$, so that grandparent's income and health become more important as the grandchild ages, and recent exposures matter more than earlier exposures. Approximately 76.51, 64.53, and 9.45 percent of grandparents survive beyond the first, second, and third test date, respectively. The mean (standard deviations) of grandchildren's test scores are 5916.61 (2726.47), 10513.63 (5528.72), and 18320.90   (13378.57) points at the first, second, and third test date, respectively. Descriptively, the test scores of grandchildren whose grandparent had died were 26531.23 points lower than the test scores of grandchildren whose grandparent had not died by the third test date. 

\subsection{Simulation Results}

From this process, we generate 2,000 samples, each containing $N=100,000$ grandparent-grandchild pairs, $i$, to simulate the relevant sampling distributions. This sample size roughly corresponds to those available in typical population registers. 

Estimation follows the four-step procedure developed in section \ref {sec:hetero}. Using the first two steps, we estimate the parameters of grandparents' time-varying characteristics, $\beta_{t'}$, between the first and third test date, $\beta_{11}$ through $\beta_{20}$. Using the third step, we estimate the contribution of the unobserved component ($U_{i2}$) to the grandparent overlap effect. In the fourth step, we estimate various grandparent overlap effects. For this illustration, we choose the characteristics of deceased grandparents during their imagined additional year of life, $C_{A_{iq3}+1}$, to equal the observed characteristics of the grandparents who die in the subsequent year, $A_{iq3}+1$, \footnote{ $\hat E\big[C_{iA_{iq_{3}}+1}\big| A_{iq_{3}}=a, q_2<A_{iq_{3}}<q_3 \big]=\frac{1}{m_{a+1}}\sum_{i\in A_{iq3}=a+1} C_{iA_{iq3}+1}$, where $m_{a+1}$ is the number of grandparents whose length of overlap is $A_{iq3}=a+1$.} although different analyst may have made different choices. 

Table 2 shows the true values and the average estimates of the three components constituting the length-specific average overlap effects on test scores at age 20, $LAOE_{20}(a)=E\big[C_{a+1}\beta_{a+1}+U_{i2}\big|A_{it}=a \big]$, for various specific lengths of overlap between the second and third test date, $10< a<20$: (i) the parameters on grandparent income and health, $\beta_{a+1,I}$ and $\beta_{a+1,H}$; (ii) the mean of the unobservable components, $E(U_{i2}|a)$; and (iii) the investigator-chosen value of $E(C_{i,a+1}|a)$, which contains the averages of the values to which we suppose grandparents' income and health would be set if grandparents' life would be extended by one year to age $a+1$. Table 2 shows that all estimates of the $LAOE_{20}(a)$s and all of their components are unbiased, and Appendix C illustrates that they are approximately normally distributed. For example, the causal effect of increasing grandparent overlap by one year among grandchildren who lose their grandparent at age $a=17$  on test scores at age 20 is estimated at $\overline{\widehat{LAOE}}_{20}(a=17)=2254.82$ points on average, virtually identical to the true (simulated) $LAOE_{20}(17)=2254.84$. 

We estimate the standard errors of the estimated $\widehat{LAOE}_{20}(a)$ with 500 bootstraps. The average of the boostrapped standard errors across 200 samples, $\overline{\widehat{SE}}(\widehat{LAOE}_{20}(a))$, is shown in the penultimate column. We evaluate the performance of bootstrapped standard errors by computing coverage rates, $CR(\widehat{CI})$, which give the proportion of simulations in which the estimated $95\%$ confidence interval includes the true $LAOE_{20}(a)$ , $\widehat{CI}=\widehat{LAOE}_{20}(a)\pm1.96*\widehat{SE}(\widehat{LAOE}_{20})$, shown in the last column. Since the coverage rates are fairly close to the nominal 0.95 level for all estimates, we conclude that the bootstrap method of inference performs well for our estimators. 

\begin{center}
  [Table 2 About Here]  
\end{center}

\begin{figure}[!htbp]
		\centering
			\begin{minipage}{1\linewidth}
		\includegraphics[width=\linewidth]{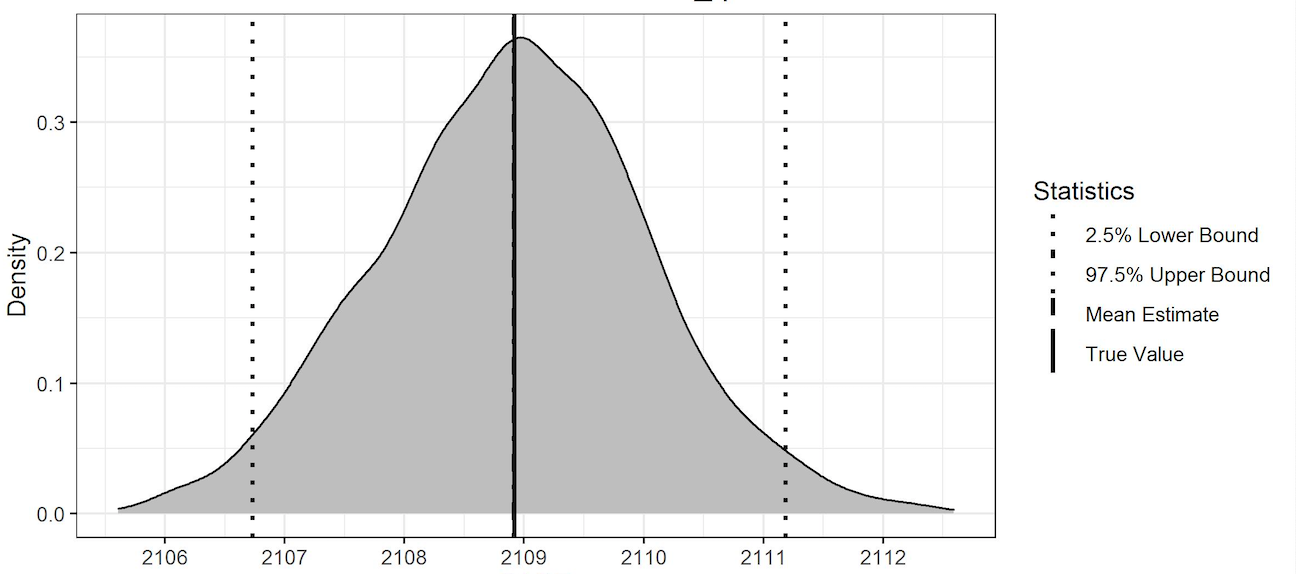}
			\caption{Sampling distribution of $\widehat{PAOE_{20}}$}
\footnotesize		
\emph{Notes:} Sample density plot of estimates for $\widehat{PAOE_{20}}$ from 2000 samples of size $N=100,000$. The dashed line indicates the sample average of the estimates, $\overline{\widehat{PAOE_{20}}}=2108.78$ ($\overline {\widehat{SE}}(\widehat{PAOE}_{20}=3.78)$, averaged from 200 sample estimates of $\widehat{SE}(\widehat{PAOE}_{20})$ from 500 bootstraps, coverage rate is 0.97), which is virtually identical to the true value, $PAOE_{20}$=2108.93.
	\label{fig:paoe}
	\end{minipage}
\end{figure}

Figure \ref{fig:paoe} shows the sampling distribution of the population-average grandparent overlap effect, $PAOE_{20}$, which is the average effect of increasing grandparent overlap by one year on grandchild test scores at age 20 among grandchildren whose grandparents died in any year between the second and third test date (when grandchildren were aged 11 to 19), $PAOE_{t}=E\big[C_{A_{it}+1}\beta_{A_{it}+1}+U_{i2}|10<A_{it}<20\big]$. The $PAOE_{20}$ is simply the average of the $LAOE_{20}(a)$s between the ages of 11 and 19, shown in Table 2, weighted by the fraction of grandparents who died at each of these ages. Since the estimates for the $LAOE_{20}(a)$s are unbiased, the estimate for the $PAOE_{20}$ is also unbiased, and the estimate is approximately normal. 

\begin{center}
    [Figure 1 About Here]
\end{center}

The overlap effects estimated in Figure \ref{fig:paoe} and Table 2 exhaust the average and length-specific grandparent overlap effects, $PAOE_t$ and $LAOE_t(a)$ that can be estimated when the second and third test scores are measured at grandchild ages 10 and 20. Notably, these effects would change if the analyst chose to explore different counterfactual values of grandparent characteristics for grandparents' additional year of life than we did here. 

Analysts could further explore effect heterogeneity along grandparent's baseline characteristics $B_i$ (which may include any grandparent fixed characteristics and time-varying observables measured before the first tests date, $q_1$ ). 

\begin{figure}[!htbp]
	
		\centering
			\begin{minipage}{1\linewidth}
		\includegraphics[width=\linewidth]{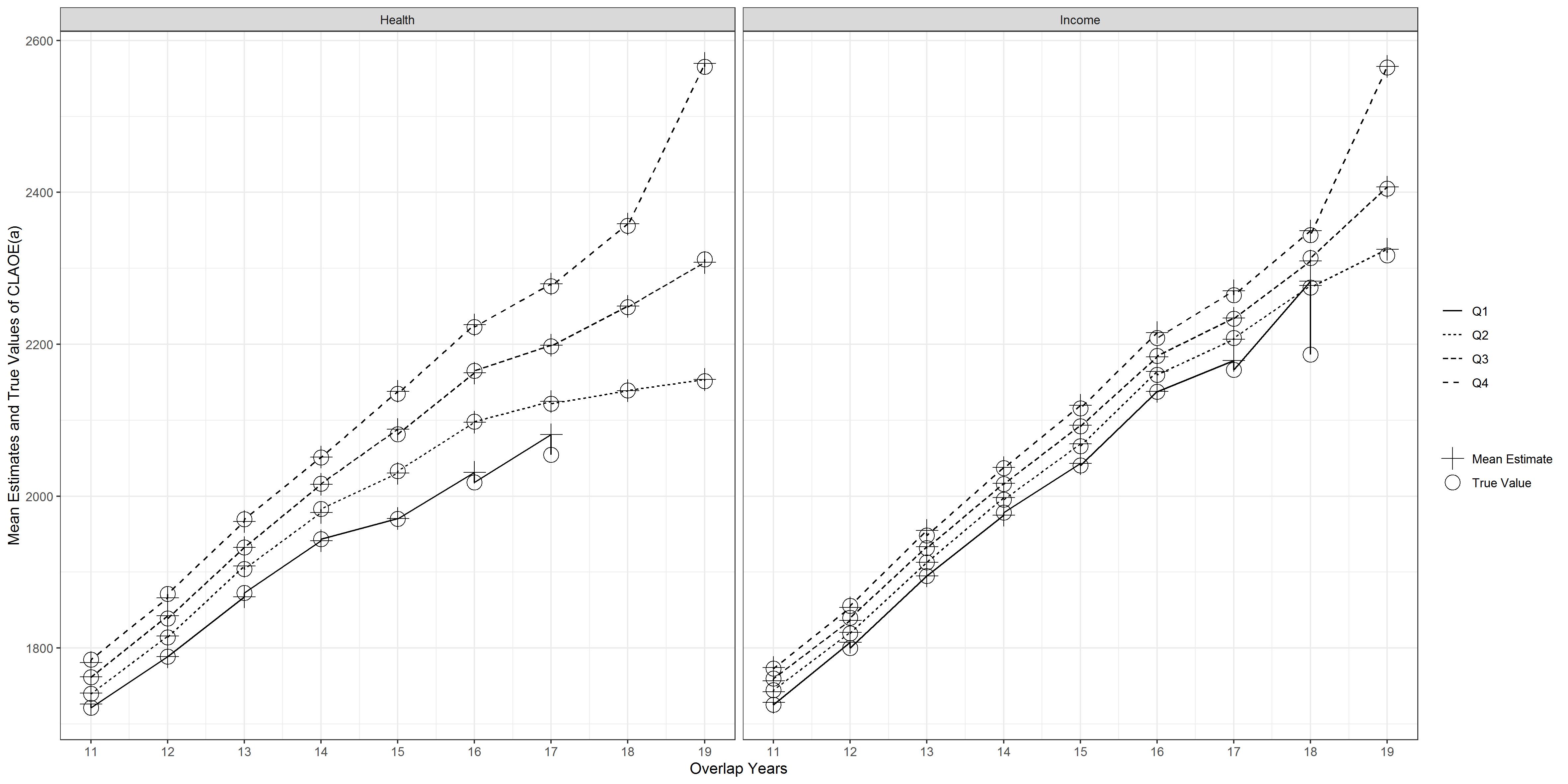}
		\caption{Subgroup mean estimates and true values of conditional length-specific grandparent overlap effects, $\overline{\widehat{CLAOE}_t^b}(a)$ and $CLAOE_t^b(a)$}
		
\footnotesize		
\emph{Notes:} Panels show the mean estimates and true values for the conditional length-specific grandparent effects defined by quartiles of grandparents' baseline health (left) and income (right) at different lengths of overlap $A_{i20}=a$, across 2000 samples of size N=100,000. These values at $A_{i20}=17$, $A_{i20}=18$ and $A_{i20}=19$ for the grandparents of the bottom quartile (Q1) of baseline income and health are incomplete or show a higher bias because most of grandparents of the lowest baseline income or health do not survive to these later stages. 

	\label{fig:subgroup}
	\end{minipage}
\end{figure}

To illustrate, we estimate the conditional grandparent overlap effects, $CPAOE_t^b$ and $CLAOE_t^b(a)$, defined by grandparent's quartiles of income and health at grandchildren's birth, by repeating the preceding analysis in samples stratified by these baseline characteristics. As  shown by Table 3, all subgroup estimates are unbiased. The population average grandparent overlap effect $CPAOE_t^b$ is greater among grandchildren whose grandparent had higher baseline income and better baseline health. \autoref{fig:subgroup} shows the trends of $CLAOE_t^b(a)$ for grandparents with different baseline health and income. The grandparent overlap effect of an additional year with grandparents who were wealthier and healthier at grandchildren's birth increase as overlap increases, while the overlap effects of grandparents with lower initial income or more health problems decrease as overlap increases.This exercise demonstrates that our approach can estimate substantially nonlinear grandparent effects. 

As before, we estimate the standard errors of the conditional grandparent overlap effects with 500 bootstraps. The estimated standard error and coverage rates of the estimated $\widehat{CPAOE_t^b}$, $\overline{\widehat{SE}}(\widehat{CPAOE}_t^b)$ and $CR(\widehat{CI})$, are shown in the last two columns of Table 3. (Standard errors and coverage rates for the $\widehat{CLAOE_t^b(a)}$ are available upon requests). Again, the bootstrap performs well, with coverage rates close to the nominal 0.95 rate.

\begin{center}
  [Table 3 About Here]

[Figure 2 About Here]
\end{center}

In sum, these simulations demonstrate that our cumulative fixed effects (CFE) approach recovers unbiased estimates of various average grandparent overlap effects, and even allows for the exploration of effect heterogeneity, in data where grandparent overlap effects vary systematically across individuals and are confounded by fixed and cumulative unobserved characteristics. With real data, these estimation options will provide sociologists with a detailed picture of different grandparent overlap effects across various durations of overlap and different grandparental characteristics.

\section{Discussion: Model Extensions}
\label{sec:discussion}

So far, we have proposed a new, multidimensional conceptualization of overlap effects, defined various causal overlap estimands and shown how they can be identified from data using a novel cumulative fixed effects estimator, relative to a flexible data generating model that conceptualizes grandparent overlap effects as the accumulation of observed and unobserved grandparental influences across the shared live course. Despite its generality, however, our model for grandparent overlap effects in \autoref{eq:dgp} naturally makes simplifying assumptions compared to an arguably more complex social reality. Because causal identification is always relative to the analyst's assumptions about the DGP, we next discuss several directions in which our DGP could fruitfully be generalized, while retaining our ability to estimate various overlap effects.  

\subsection{More Flexible Parametric Specifications}
\label{sec:gap_varying}

First, one could consider a more flexible parameterization for the effects of grandparental characteristics on grandchild outcomes. In \autoref{eq:dgp}, the effect of each grandparental characteristic, $C_{it'}$, depends only on the age, $t'$, at which the characteristics is experienced by the grandchild, but not on the age, $t$, at which the outcome is measured. This specification prevents the effects of exposures experienced at a particular age to change (increase or decrease) over time. For example, having had a healthy rather than sick grandparent at age 10 is stipulated to have the same effect, $\beta_{10}$, on grandchild's test scores at age 10 as at age 20, $Y_{i10}$ and $Y_{i20}$,  

To relax this restriction, one could allow the effects of grandparental characteristics, $C_{it'}$, to depend both on the age at which the characteristic is experienced, $t'$, and on the time that has elapsed between exposure and the measurement of the outcome $Y_{it}$ (\citealt{todd2003specification}), so that each $\beta_{it'}$ is replaced by a series of $\beta_{t}^{gap}$, $gap=t-t'$. Hence, the effect of experiencing a healthy rather than a sick grandparent at age 10 on test scores at age 10 could differ from the effect on test scores at age 20, $\beta_{10}^0\neq \beta_{10}^{10}$.

Relaxing the DGP in this manner may increase sociological realism, but it also comes at a cost. Whereas the parameters, $\beta_{t'}$ from the proposed DGP in \autoref{eq:dgp} can be identified between the first and third test date, the more flexible parameters $\beta_{t'}^{gap}$ can only be identified between the second and third test date (see Appendix D for details). This cost, however, is arguably minor, since even in the more restrictive DGP of \autoref{eq:dgp}, the grandparent overlap effects, $LAOE_{t}(a)$ and $PAOE_t$, which are the primary focus of our analysis, are only identified between the second and third test date, and this is still the case in this more general DGP. 

Apart from the more flexible parameter specification, the model can be flexibly extended to accommodate other parametric forms of interactions. A first scenario involves the interaction terms of observables, such as different grandparent observed time-varying characteristics (e.g. income and health) and grandparent characteristics at different time points $Inc_{it}$ and $Inc_{it-1}$, and the latter is useful to account for the cumulative advantage of having stably-high-income grandparents. A second scenario involves more flexible specification of the unobserved overlap effect, say, from $U_{i2}*A_{it}$ to $U_{i2}(A^2+0.5A)$. In fact, applicants may follow the same four-step procedures of CFE to rescale and eliminate the term of $U_{i2}g(A_{it})$ as long as the parametric form of $g(A_{it})$ is known and correctly specified. Similarly, the model can be extended to address the scenario where the unobserved fixed confounder is interactive with both the length of overlap and observed characteristics, taking the form of $U_{i3}*A_{it}*C_{it}$. Since $A_{it}*C_{it}$ is observed, applicants could add a triple difference step to rescale and elimiate $U_{i2}g(A_{it}, C_{it})$ and achieve the identification of the parameters $\beta_{t'}$. This requires a slight modification of the estimation (i.e. a triple difference) and at least four waves of individual panels.

\subsection{Death Effects}
\label{sec:death effct}

Second, one could elaborate on the role of grandparent's death in determining grandchildren's test scores. The DGP of \autoref{eq:dgp} describes a model in which grandparent-overlap effects originate exclusively from the grandchild's cumulative exposure to grandparental characteristics while the grandparent is alive. The death of the grandparent in this model simply marks the end of grandparental exposure, but does not exert an effect on grandchild test scores by itself. One could reasonably explore more general models, in which grandparent's death does affect grandchildren's test scores directly, for example, via inheritance (\citealt{hallsten2017grand}) or grief (\citealt{silverman2000never}). 

Theorizing such death effects would be interesting, because they might trade-off in complex fashion with the grandparent overlap effect understood, as before, as the effect of cumulative exposure. Specifically, dying one year later would increase the grandchild's exposure to the grandparent (likely a positive influence on the grandchild), but plausibly would also reduce the amount of the inheritance that the grandchild receives (likely a negative effect); then again, dying one year later would mean that more of whatever inheritance the grandchild may have received is left over at the time of the fixed subsequent test date (likely a positive effect), but grief is more acute, too (likely a negative effect).  Hence, the total effect of cumulative grandparental exposure plus the postponement of death resulting from an additional year with the grandparent on the grandchild's test scores could be positive or negative. 

Such considerations could be explored in elaborated models, such as that of \autoref{eq:dgp_af}, which adds death effects via inheritance to \autoref{eq:dgp},
\begin{align}
&Y_{it}=\beta_{0}+\sum_{t'=1}^{A_{it}}C_{it'}\beta_{t'}
+D_{it}N_{i}\gamma +U_{i1}+U_{i2}A_{it}+\epsilon '_{it}
\label{eq:dgp_af}
\end{align}
where $D_{it}=1$ is an indicator for grandparent's death, $=0$ if alive; $N_i$ measures the amount of the inheritance and grief; and $\gamma$ is the effect of the inheritance on grandchild test scores.\footnote{To allow that the effect of the inheritance diminishes over time, one could define $N_{it}$ as the present value of the inheritance (suitably defined) or permit $\gamma_{t-d_i}$ to vary freely with time since grandparent's death.}

When $N_i$ is observed, our estimation approach can still identify grandparent overlap effects by adjusting for $N_i$ explicitly. The interpretation of the estimates, however, would subtly change.  Rather than estimating the total effect of postponing a grandparent's death, our estimation approach would now identify the controlled direct effect of grandparent overlap, net of fixing the amount of inheritance and grief. Hence, the $PAOE_{t}$ would no longer identify $E_{\bar C_t,A_{it}}\big[E\big[Y_{it}(a+1)-Y_{it}(a)| q_2<A_{it}<t\big]\big]$ but $E_{\bar C_t,A_{it},N_i}\Big[E\big[Y_{it}(a+1, N_i)-Y_{it}(a, N_i)|q_2<A_{it}<t\big]\Big]$. 

When $N_{it}$ is unobserved and $\gamma \neq 0$, our approach cannot identify the model in \autoref{eq:dgp_af}. This is because grandparent's death, $D_{it}$, is not random, but is a function of the history of past grandparent characteristics, $\bar C_{it'}$, and the fixed unobservables, $U_i$. When the term $D_{it}N_i\gamma$ is non-zero, it cannot be differenced out by the double differencing of estimation steps 1 and 2 above, so that a function of the unobserved confounders $U_i$ would remain in the error term of the regression.

Since few grandparents leave significant bequests (\citealt{wolff2014inheritances}), and those who do often divide the bequest over multiple beneficiaries, it may be permissible to ignore the issue of inheritances. Once the Pandora's box of death effects is opened, however, sociologists should at least discuss the complications arising from typically unmeasured consequences of bereavement, such as grief, on grandchild educational test scores. Further work should explore the likely magnitude of bias from ignoring death effects--a possibility that would certainly benefit from focused sociological theorizing about the consequences of grandparental death and their temporal articulation with respect to specific grandchild outcomes.   

\subsection{Multiple Grandparents}
\label{sec:ext_jointGP}

 The model of \autoref{eq:dgp} refers to a single focal grandparent. One could expand the model to include multiple grandparents in two ways. First, one could keep the focus on estimating grandparent overlap effect due to a single focal grandparent. In this case, \autoref{eq:dgp} could be extended to include grandparent's spouse's characteristics as potential time-varying confounders. For example, the focal grandparent's spouse's death could affect the focal grandparent's survival through widowhood effects (\citealt{elwert2008wives}). When the grandparent's spouse's characteristics are potential time-varying confounders, they should be included for in the model. Identification, however, would be compromised if spouse's characteristics are not only confounders of the effect of focal grandparent's characteristics and grandchild outcomes, but also mediators of the effect of focal grandparent's prior characteristics on grandchild outcomes. In this event, the option discussed next would offer a solution.   

 Second, one could move away from modelling the overlap effect due to a single focal grandparent and instead model the joint overlap effect of multiple grandparents on a focal grandchild. One straightforward way to accomplish such a joint model would be to regard the grandmother's characteristics and vital status as the grandfather's spousal characteristics in $C_{it}$, and to allow the joint overlap effect to accumulate across the grandfather's overlap, or, alternatively across the grandmother's overlap. Such an approach would model each grandparent separately while exploiting the information of his or her spouse. Another approach would be to aggregate $C_{it}$ and $A_{it}$ over each lineage of grandparents, using either their minimum, mean, or maximum survival to define $A_{it}$. The DGPs including joint grandparent time-varying confounders would be analogous to the form of \autoref{eq:dgp} with only slight adjustments to notation, and analysts should be careful in choosing clustered standard error to address problems of inference due to non-independence.

\section{Conclusion}
\label{sec:conclusion}

The paper addresses the conceptual and methodological issues with overlap effects, when interest lies in the effect of changing the duration of exposure of one individual (or complex entity) to another individual, group, or environment. Some important sociological questions involve, for instance, the estimation of effects of father’s overlap, grandparent overlap, teacher’s overlap and neighborhood overlap.

The idea of the causal estimand of changing time exposure is, of course, no particular mystery. Conventional framework of causal inference, however, has regarded the overlap effects as the causal effects of an intervention on the length of overlapping time, a unidimensional treatment effect. Such defined overlap effect violates the notion of a ``well-defined” hypothetical interventions in the potential outcomes framework (Vanderweele 2018), since the state or event of X=x in itself is insufficient to fix the value or the distribution of $Y_{x}$ for each individual in the population under consideration. This complication occurs since the overlap is a non-manipulative treatment which leads to changes in the potential outcomes not due to an isolated intervention or human action, but rather because of the entailed changes in the state of the universe, i.e. the length of accumulation of the effect of the contextual characteristics that one is exposed to, which imposes a violation to the ``no multiple versions of the treatment” condition under SUTVA (Rubin 1979).  

Our framework offers a way out by newly operationalizing the overlap effects as multidimensional treatments. Formally, we define overlap effects as multidimensional effects which capture both the “duration” of overlap and the “context” of overlap, i.e. the effect of extending an individual’s exposure to all of the contextual observed and unobserved characteristics. The framework acknowledges the social reality that the potential outcomes of, say, the grandchild’s cognitive outcome, can be flexibility dependent on the grandparent characteristics during the extended period of grandparent overlap, and thus encompasses numinous treatment heterogeneity. 

Substantively, we next proposed a highly flexible formal model for overlap effects on the individual's outcomes  that explicitly relates the outcomes to the history of the contextual observed and unobserved characteristics across the shared time exposure, and we subsequently discussed several ways in which this model could further be generalized. 

Estimating such multidimensional effects, however, is difficult, since none of the usual tools of causal inference in sociology realistically apply: Randomized experiments are either practically infeasible (e.g. prolonging grandparents' life) or unethical (killing grandparents before their time). Conventional regression and matching estimators suffer the obvious threat of unobserved confounding. Relevant and valid instrumental variables are hard to come by. And conventional fixed effects estimators are inappropriate, because overlap effects plausibly vary with the unobservables. 

Practically, we finally proposed a new cumulative fixed effects (CFE) estimator that can recover our proposed suite of overlap effects relative to the proposed data-generating model from panel data under the conventional strict exogeneity assumption (essentially: no unobserved time-varying confounders). Whenever duration of exposure to a complex entity cannot be estimated directly, our approach allows analysts to disaggregate the entity into its observed and unobserved characteristics in order to ``reassemble" the effect of exposure to the entity from the effects of exposure to the entity's characteristics.

By adopting the multidimensional framework, analysts  are prompted to clarify the characteristics of the entity during the hypothetical extension of overlap, hence forcing greater clarity and realism of otherwise implicit counterfactuals. In this process, important simplification assumptions of a unidimensional overlap framwork can get relaxed, such as no time-varying confounders and no interactive fixed effect of overlap.  
 
\newpage

\theendnotes

\newpage
\bibliographystyle{asr}

\bibliography{method_ref}

\newpage

\section*{Tables}

\begin{table}[H] 
	\centering
		\caption{Notation}
	\renewcommand{\arraystretch}{1.2}
	\begin{tabular}{rlrrrrrrr}
		\hline
		\hline
		$i$ & a grandchild-grandparent pair \\
		\hline
		 $t$, $t'$ & grandchild's age at the outcome, \\ & running index of grandchild's age\\
		 \hline
		 $Y_{it}$ &  outcome of individual grandchild $i$ at time point $t$ \\
		\hline
		$A_{it}$ &  duration of grandparent overlap of grandchild $i$ at grandchild's age $t$ \\
		\hline
		$\mathscr{A}_{it}$ &  multidimensional treatment of overlap of grandchild $i$ at grandchild's age $t$ \\
		\hline
		$d_i$ &  grandchild's age at grandparent's death \\
		\hline
		$B_i$ & the vector of grandparent baseline characteristics\\
		\hline
		$C_{it'}$ & the vector of time-varying grandparent characteristics at $t'$ \\
		\hline
		$\bar C_{it'}$ & history (i.e. complete sequence) of grandparent characteristics, $C_{it'}$, \\
		& from $i$'s birth to age $t'$ \\
		\hline
		$U=(U_{i1}, U_{i2})$ & effect of grandparent unobservables \\ & $U_{i1}$  comprises the fixed effect that affect $Y_{it}$ marginally. \\& $U_{i2}$ comprises the effect of unobservables that affect $Y_{it}$ in interaction with $A_{it}$.  \\
		\hline
		$\epsilon_{it}$ & $i$'s exogeneous idiosyncratic time-varying unobservables (error term)\\
		\hline
		$q_1$, $q_2$ and $q_3$ & grandchild's age at the three outcome assessments, $q_1<q_2<q_3$\\
		\hline
		$D_{it'}$ & time-varying dummy indicator for the death of $i$'s grandparent at $t'$\\
		\hline
		$I_{it'}$ & grandparent income at $t'$ \\
		\hline
		$H_{it'}$ & grandparent health at $t'$ \\
		\hline
		$P_{it'}$ &  time-varying parent characteristics at $t'$ \\
		\hline
		$N_i$ & amount of inheritance $i$ would receive\\
		\hline
		\hline
	
	\end{tabular}
\end{table}

\begin{landscape}
\begin{table}[H]
\begin{threeparttable} 
	\renewcommand{\arraystretch}{1.2}
\caption{Elements of the four steps in estimating $LAOE_{20}(a)$}
\small
\label{Tab:est}
\begin{tabular}{ p{0.3cm}|p{0.9cm}|p{0.9cm}| p{1cm}|p{1cm}|p{1cm}|p{1.2cm}|p{1.3cm}|p{1.3cm}|p{1.9cm}|p{1.9cm}|p{2.7cm} |p{1.2cm}}
\hline
\hline
$a$	& $\beta_{a+1,I}$ &	$\overline{\widehat{\beta}}_{a+1I}$	&	$\beta_{a+1,H}$	&$\overline{\widehat{\beta}}_{a+1H}$	&	$\bar I_{ia+1}$	&	$\bar H_{ia+1}$	&	$E(U_{i2}|a)$	&	$\overline{\widehat{E}}(U_{i2}|a)$	&	$LAOE_{20}(a)$ &  $\overline{\widehat{LAOE}}_{20}(a)$ & $\overline{\widehat{SE}}(\widehat{LAOE}_{20}(a))$ & $CR(\widehat{CI})$ \\
\hline												
\hline							11	&	10.5	&	10.5	&	10.5	&	10.5	&	261.12	&	-93.5	&	-9.21	&	-9.47	&	1750.84	&	1750.85	&	4.33	&	0.97	\\
12	&	11	&	11	&	11	&	11	&	285.94	&	-119.91	&	4.04	&	3.78	&	1830.3	&	1830.3	&	4.45	&	0.96	\\
13	&	11.5	&	11.5	&	11.5	&	11.5	&	314.29	&	-147.9	&	17.02	&	16.76	&	1930.48	&	1930.47	&	5.39	&	0.92	\\
14	&	12	&	12	&	12	&	12	&	343.97	&	-178.45	&	33.05	&	32.78	&	2019.23	&	2019.25	&	5.84	&	0.93	\\
15	&	12.5	&	12.5	&	12.5	&	12.5	&	375.4	&	-211.17	&	49.02	&	48.75	&	2101.98	&	2101.96	&	6.88	&	0.95	\\
16	&	13	&	13	&	13	&	13	&	410.16	&	-246.07	&	65.78	&	65.51	&	2198.97	&	2198.99	&	7.99	&	0.97	\\
17	&	13.5	&	13.5	&	13.5	&	13.5	&	444.31	&	-283.68	&	86.4	&	86.12	&	2254.84	&	2254.82	&	8.63	&	0.98	\\
18	&	14	&	14	&	14	&	14	&	482.87	&	-323.15	&	104.61	&	104.33	&	2340.67	&	2340.68	&	11.24	&	0.94	\\
19	&	14.5	&	14.5	&	14.5	&	14.5	&	534.12	&	-367.03	&	128.47	&	128.18	&	2551.38	&	2551.35	&	17.43	&	0.98	\\

\hline
\end{tabular}
    \begin{tablenotes}
      \footnotesize
      \item \emph{Notes:} $a$ is the duration of overlap at $t=20$, $A_{i20}=a$.  $\beta_{a+1,I}$ and $\beta_{a+1,H}$ indicate the true values of the parameters of grandparent income and health at the extended overlap stage, $a+1$. $\overline{\widehat{\beta}}_{a+1,I}$ and $\overline{\widehat{\beta}}_{a+1,H}$ indicate the sample average of the estimates of the parameters of grandparent income and health at $a+1$. $\bar I_{ia+1}$ and $\bar H_{ia+1}$ are the average grandparent income and health at $a+1$ for those who die at this wave, which are predictions of the grandparent income and health at $a+1$ if those who die at $a$ survive to $a+1$. $E(U_{i2}|a)$ is the true values of the unobservable at different a.  $\overline{\widehat{E}}(U_{i2}|a)$ is its sample averages. 	$LAOE_{20}(a)$ are the true values of length-specific grandparent overlap effect at different a. $\overline{\widehat{LAOE}}_{20}(a)$ is the estimated sample average. $\overline{\widehat{SE}}(\widehat{LAOE}_{20}(a))$ is the average of 200 sample estimates of standard errors, each obtained from 500 bootstraps. $CR(\widehat{CI})$ (coverage rate of $\widehat{CI}$) is the proportions that the true values of $LAOE_{20}(a)$ are covered by the 200 estimated 95\% confidence intervals, each obtained from 500 bootstraps.
    \end{tablenotes}
\end{threeparttable}
\end{table}
\end{landscape}

\begin{table} [H]
\caption{True values and mean subgroup estimates of conditional population average overlap effects}
\begin{threeparttable} 
	\renewcommand{\arraystretch}{1.2}
\begin{tabular}{p{3cm}|p{2.5cm}|p{2.5cm}|p{2.5cm}|p{2.5cm}|p{2.5cm}|p{2.5cm}}
\hline
\hline
b 	& $CPAOE_t^b$ & $\overline{\widehat{CPAOE}_t^b}$ & $\overline{\widehat{SE}}(\widehat{CPAOE}_t^b)$ & $CR(\widehat{CI})$\\
\hline												
\hline
$H_{i1}$:  Q4 (best) 	&	2332.05	&	2332.05	&	6.74	&	0.94	
\\									
$H_{i1}$: Q3 	&	2055.39	&	2055.39	&	3.28	&	0.94	
\\									
$H_{i1}$: Q2 	&	1869.11	&	1869.11	&	2.72	&	0.96	
\\									
$H_{i1}$: Q1 (worst) 	&	1763.21	&	1763.22	&	9.46	&	0.97	\\
\hline 									
$I_{i1}$: Q4 (highest)	&	2299.17	&	2299.17	&	6.62	&	0.95	
\\									
$I_{i1}$: Q3 	&	2070.38	&	2070.38	&	3.26	&	0.94	\\
$I_{i1}$: Q2 	&	1885.39	&	1885.39	&	2.63	&	0.97	\\
$I_{i1}$: Q1 (lowest) 	&	1777.77	&	1777.77	&	8.52	&	0.98	\\

\hline
\end{tabular}
 \begin{tablenotes}
      \footnotesize
      \item \emph{Notes:} $b$ indicates the baseline grandparent characteristics, $B_i=b$, that define the subgroups across which the conditional effects are averaged: quartiles of grandparent health, $H_{i1}$, and grandparent income, $I_{i1}$. $CPAOE_t^b$ indicates the true value of the conditional population average overlap effects at specific levels of b; $\overline{\widehat{CPAOE}_t^b}$ is the average of the estimates across $200$ simulations with $N=100,000$ observations. All estimates are unbiased.  $SE(\widehat{CPAOE}_t^b)$ indicates the standard errors of $\widehat{CPAOE}_t^b$ from 500 bootstraps. $\overline{\widehat{SE}}(\widehat{CPAOE}_t^b)$ is the average of 200 sample estimates of the standard errors of $\widehat{CPAOE}_t^b$, each obtained from 500 bootstraps. $CR(\widehat{CI})$ (coverage rate of $\widehat{CI}$) is the proportions that the true values of $CPAOE_t^b$ are covered by the 200 estimated 95\% confidence intervals, each obtained from 500 bootstraps.
      
    \end{tablenotes}
\end{threeparttable}
\label{Tab:subsample}
\end{table}

\newpage 
\section*{Figures}

\newpage
\section*{Appendix}

\subsection*{A. Generalized Methods of Moment Estimation Procedure}

The four-step estimation approach of CFE has the advantage of mapping each components of the estimand to the statistical procedures clearly, which illustrates our ideas of identification. But this is not the only available estimation procedure. Analysts can recover the estimands with a generalized methods of moment estimation procedure (GMM). The GMM procedure can be more accessible because analysts may achieve the estimands by specifying the required moment conditions directly using the standard Stata packages (e.g. the ``gmm" command), which requires minimum programming.  

The following section includes the required moment conditions for recovering the estimands. We will also show how these moment conditions can be derived from the equations in the main text of the four-step CFE estimator. 

\subsubsection*{Moment Conditions for Estimating $PAOE_t$ and $LAOE_t(a)$}

Two moment conditions are required to estimate each one of $PAOE_t$ and $LAOE_t(a)$. 

The first moment condition for estimating $PAOE_t$ aims at estimating $\beta_t$,

\begin{align}
E(\frac{\bigtriangleup_{q_2q_1}Y_{i}}{\bigtriangleup_{q_2q_1}A_{i}}-\frac{\bigtriangleup_{q_3q_2}Y_{i}}{\bigtriangleup_{q_3q_2}A_{i}})-&E(\frac{\sum_{t'=q_1+1}^{A_{iq_2}}C_{it'}\beta_t}{\bigtriangleup_{q_2q_1}A_{i}}-\frac{\sum_{t'=q_2+1}^{A_{iq_3}}C_{it'}\beta_t}{\bigtriangleup_{q_3q_2}A_{i}})=0
\label{eq:gmm1_AOE}
\end{align}

which can be derived from equation \ref{eq:first_diff2} and equation \ref{eq:assumption} (the assumption of strict exogeneity) which posits that $\frac{\bigtriangleup_{q_2q_1}\epsilon_{i}}{\bigtriangleup_{q_2q_1}A_{i}}-\frac{\bigtriangleup_{q_3q_2}\epsilon_{i}}{\bigtriangleup_{q_3q_2}A_{i}}=0$. 

The second moment condition for $PAOE_t$ is

\begin{align}
&\frac{E\big(\bigtriangleup_{q_3q_2}Y_{i}-\sum_{t'=q_2+1}^{A_{iq_3}}C_{it'}\beta_t \big|A_{iq_3}, \bar C_{iA_{iq_3}}\big)}{\bigtriangleup_{q_3q_2}A_{i}} -PAOE_t+E\big[C_{iA_{it}+1}\beta_{A_{it}+1}| A_{it}<t\big]=0,
\end{align}

which can be derived from the definition of $PAOE_t= E[C_{iA_{it}+1}\beta_{A_{it}+1}+U_{i2}|A_{it}<t]$ (equation \ref{eq:PAOE}) and the third step of recovering the effect of the unobservables, $E(U_{i2}|A_{it}<t)$ (equation \ref{eq:CEres_final}). 

As mentioned in the main text, analysts can recover the $LAOE_t(a)$ for $q_2<a<q_3$. The moment conditions for estimating $LAOE_t(a)$ are similar to the conditions for $PAOE_t$ but are different in that they are $A_{it}=a$ specific. They are

\begin{align}
E(\frac{\bigtriangleup_{q_2q_1}Y_{i}}{\bigtriangleup_{q_2q_1}A_{i}}-\frac{\bigtriangleup_{q_3q_2}Y_{i}}{\bigtriangleup_{q_3q_2}A_{i}})-&E(\frac{\sum_{t'=q_1+1}^{A_{iq_2}}C_{it'}\beta_t}{\bigtriangleup_{q_2q_1}A_{i}}-\frac{\sum_{t'=q_2+1}^{A_{iq3}=a}C_{it'}\beta_t}{\bigtriangleup_{q_3q_2}A_{i}})=0
\label{eq:gmm1_AOE}
\end{align}

and 

\begin{align}
&\frac{E\big(\bigtriangleup_{q_3q_2}Y_{i}-\sum_{t'=q_2+1}^{A_{iq_3}=a}C_{it'}\beta_t \big|A_{iq_3}=a, \bar C_{ia}\big)}{\bigtriangleup_{q_3q_2}A_{i}} -LAOE_t(a)+E\big[C_{iA_{it}+1}\beta_{A_{it}+1}| A_{it}=a<t \big]=0
\end{align}

In application using the ``gmm" package of STATA, analysts are adviced to conduct subsample analysis based on $A_{it}=a$ to estimate $LAOE_t(a)$.

\subsection*{B. Parametric Form of the Simulation}

This appendix gives the parametric form of the data generating processes for the simulations reported in section 6. The process involves seven steps. For each of $N$ disjunct grandparent-grandchild pairs, $i$:

1. Draw exogenous grandparent's fixed effects $U_{i} \sim N(10,100)$, and set $U_i=U_{i1}=U_{i2}$. 

2. Generate grandparent's age at grandchild's birth $Age_{i1}$ as a function of $U_i$, so that $Age_{i1}$ increases with $U_i$.
\begin{align*}
&Age_{i1}=percentile(U_{i})*10+50\\
\end{align*}
3. Generate baseline values for grandparent income (and good health) as increasing (decreasing) functions of grandparent's baseline age, $Age_{i1}$:
\begin{align*}
& I_{i1} \sim N(100+Age_{i1},1),\quad H_{i1}\sim N(100-Age_{i1},1)\\
\end{align*}
4. Generate the time series of grandparent income and health,
$C_{it'>1}=\{I_{it'>1},H_{it'>1}\}$, as functions of (i) their initial values $I_{i1}$ and  $H_{i1}$ (ii) each other's most recent values, $I_{it'-1}$ and $H_{it'-1}$, (iii) grandparent's fixed characteristics, $U_{i}$, and (iv) their own time trends, $t'_I$ and $t'_H$:

\begin{align*}
&For\, t'>1, I_{it'}=0.95I_{it'-1}+0.035H_{it'-1}+0.03U_{i}+0.02I_{i1}t'_I+e_{Iit}\\
&For \,t'>1, H_{it'}=0.85H_{it'-1}+0.005
I_{it'-1}-0.02U_{i}+(-0.004t'_H-0.006{t'_H}^2)*H_{i1}+e_{Hit}\\
&where \quad e_{Iit}\sim N(0,1)  \quad \independent \quad e_{Hit}\sim N(0,1)\\
\end{align*}

5. Generate a sequence of 20 binary indicators for grandparent's survival at grandchild's age $t'$, $D_{it'}$, so that grandparent's survival ($D_{it'}=0$) at any $t'$ correlates positively with their income and good health at $t'-1$ and with their fixed characteristics, $U_{i}$:
\[
D_{it'}=\begin{cases}
               1 (dead), \quad if\quad  I_{it'-1}+H_{it'-1}+U_{i}+e_{it}<100\\
               0(alive),\quad if\quad I_{it'-1}+H_{it'-1}+U_{i}-0.5*t^2+10 \geq 100
            \end{cases}
            \]
It follows that the grandchild's age at the grandparent's death is $d_i=20-\sum_{t'=1}^{20}{ D_{it'}}$. 

6. Generate the duration of overlap at age $t$ as $A_{it}=\min(d_i,t)$.

7. Generate grandchild's test scores, $Y_{it}$ at three test dates, $q_1=6$, $q_2=10$ and $q_3=20$, according to the reduced-form DGP of \autoref{eq:dgp}, where grandparents' observed characteristics measured at times $t'$ affect grandchildren's test scores with parameters $\beta_{t'I}=\beta_{t'H}=5+0.5*(t'-1)$. Since the CFE estimator requires that grandparents survive beyond the second test date, the outcomes are generated by

\begin{align*}
Y_{i6}=&5*I_{i1}+5*H_{i1}+...+7.5*I_{i6}+7.5*H_{i6}+U_{i1}+U_{i2}A_{i6}+\epsilon_{i6}\\\nonumber
 Y_{i10}=&5*I_{i1}+5*H_{i1}+...+9.5*I_{i10}+9.5*H_{i10}+U_{i1}+U_{i2}A_{i10}+\epsilon_{i10}\\\nonumber
 Y_{i20}=&5*I_{i1}+5*H_{i1}+...+(5+0.5*(A_{i20}-1))*I_{iA_{i20}}\\\nonumber
&+(5+0.5*(A_{i20}-1))*H_{iA_{i20}}+U_{i1}+U_{i2}A_{i20}+\epsilon_{i20}\\\nonumber
\end{align*}

\subsection*{C. Simulation Results: Sampling Distributions}

This appendix shows the simulated sampling distributions for selected $\beta_{t'}$, for $U_{i2}$, and for the $LAOE_{20}$ to illustrate that all estimates are unbiased and approximately normally distributed. 

Figure B1 shows the sampling distributions of the coefficients for the effects of grandparent's health and income at grandchildren's age $t'=18$, $\beta_{18I}$ and $\beta_{18H}$. The sampling distributions for the effects of income and health experienced at the remaining ages are similar, available upon request.

\begin{figure}[!htbp]	

	\begin{minipage}{0.45\linewidth}
		\centering
		\includegraphics[width=9cm, height=5cm]{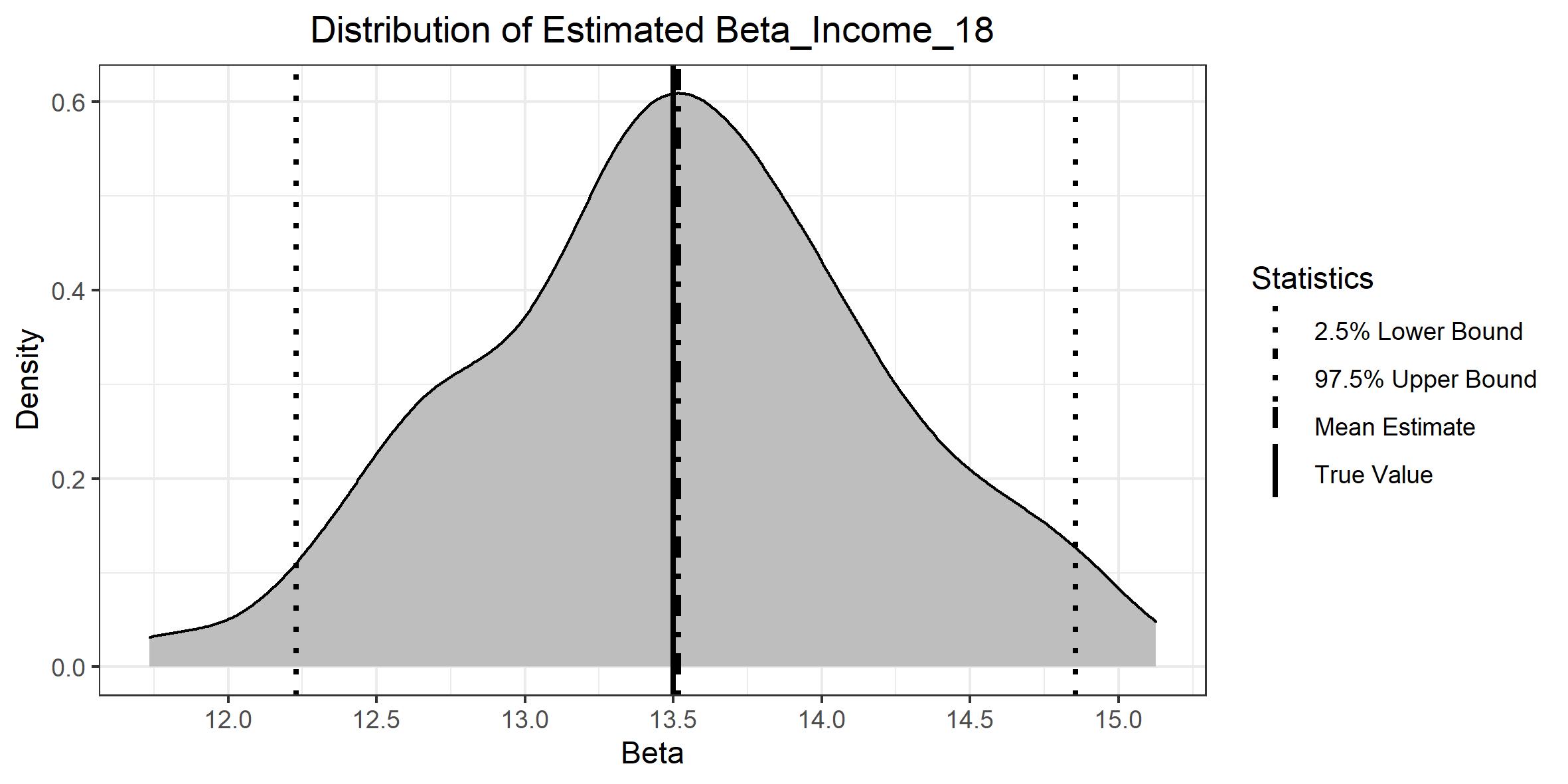}
		
	\end{minipage}
	\hfill
	\begin{minipage}{0.45\linewidth}
		\centering
		\includegraphics[width=9cm, height=5cm]{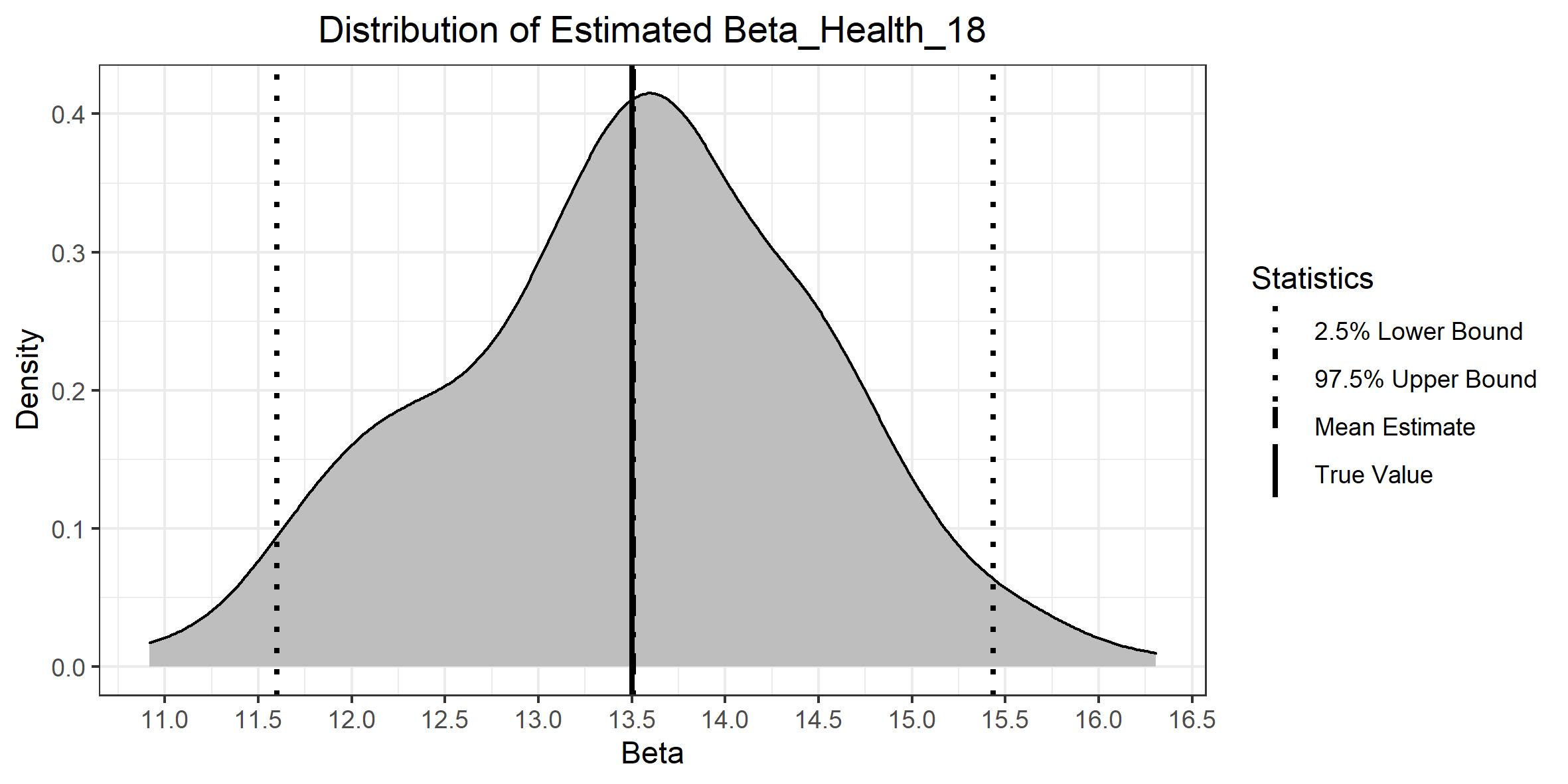}
	\end{minipage}
	\caption*{Figure B1: Sampling distributions for  $\hat \beta_{18} $} \label{fig:hist-para}
	
\footnotesize
\emph{Notes:}	The left and right panels show the sampling distributions of the coefficients for grandparent income, $\hat\beta_{18I}$, and grandparent health, $\hat\beta_{18H}$, respectively, experienced at grandchildren's age 18 on test scores at age $20$ from 2000 simulated samples of size $N=100,000$.  The means of the estimates equal the population parameters, $\hat\beta_{18}=\beta_{18}$=13.5.

\end{figure}

 Figure B2 shows the sampling distribution of the unobserved component $\widehat{E}(U_{i2}|A_{i20}=17)$ of the $LAOE_{20}(17)$. The sampling distributions at the remaining ages are similar, available upon request.

\begin{figure}[!htbp]
		\centering
			\begin{minipage}{0.8\linewidth}
		\includegraphics[width=\linewidth]{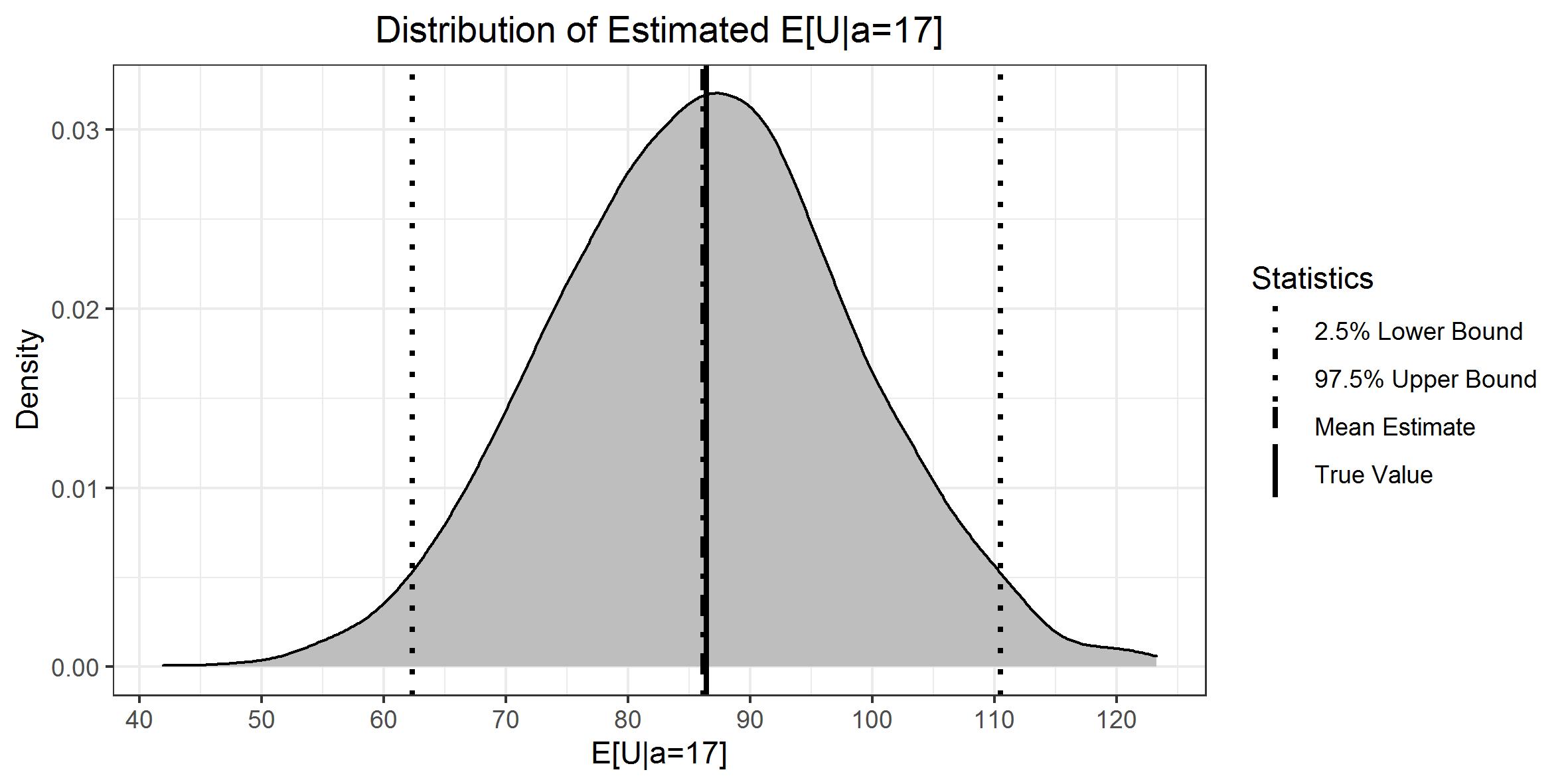}
			\caption*{Figure B2: Sampling distribution of the estimated unobservable component $\widehat{E}(U_{i2}|a=17)$}
				\label{fig:U2}
	\footnotesize
		\emph{Notes:} This graph shows the sampling distribution of the estimated unobservable component, $\widehat{E}(U_{i2}|a=17)$, for the $LAOE_{20}(17)$ from 2000 simulated samples of size $N=100,000$. The mean of the estimates, $\overline{\widehat{E}}(U_{i2}|a=17)=86.12$ is approximately equal to the true value $E(U_{i2}|a=17)$=86.40. 
		\end{minipage}
\end{figure}

Figure B3 shows the sampling distribution for the length-specific average overlap effect on grandchild test scores at age 20 for grandchildren who experienced 17 years of overlap, $LAOE_{20}(17)$. The sampling distributions at the $LAOE_{20}(a)$ at other ages, $a$, are similar, available upon request. All estimates are approximately normally distributed and unbiased.

\begin{figure}[H]
	
		\centering
			\begin{minipage}{0.8\linewidth}
		\includegraphics[width=\linewidth]{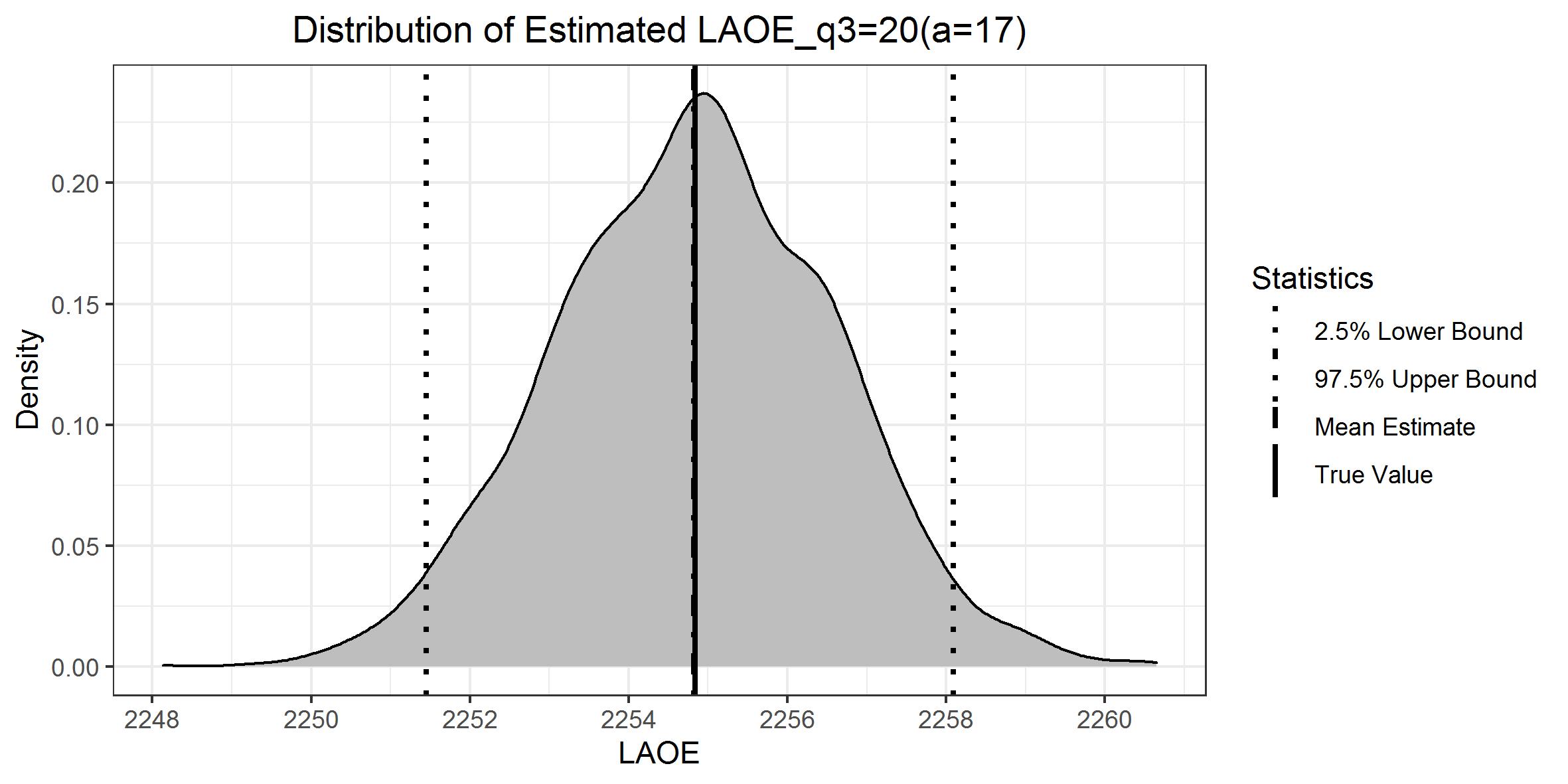}
		\caption*{Figure B3: Sampling distribution of $\widehat{LAOE}_{20}(a=17)$ }
\footnotesize		
\emph{Notes:} The sample density plot is drawn from 2000 samples. The average of the estimates, $\overline{\widehat{LAOE}}_{20}(a=17)$=2254.82 equals the true value of $LAOE_{20}(a=17)=2254.84$. 
	\label{fig:laoe}
	\end{minipage}
\end{figure}

\subsection*{D: Gap-specific Parameters, $\beta_{t'}^{gap}$}

 This appendix elaborates the proposal of section 7.1 to permit gap-specific parameters, i.e., effects, $\beta_{t'}^{gap}$, of grandparental exposures, $C_{it'}$, that vary (increase or decay) with the time elapsed between the exposure at age $t'$ and the outcome at age $t$, $gap=t-t'$. Specifically, assuming that we have three consecutive measures of the outcome, we show that identification of the gap-specific parameters and the resulting overlap effects remains feasible in the interval between the second and third outcome measurement.

As before, assume a sample in which all grandparents survive until at least the second test date, $A_{iq_3} \geq q_2$. The DGP with gap-varying parameters, $\beta_{t'}^{gap}$, at the three test date is then given by: 
\begin{align}
&Y_{iq1}=C_{i1}\beta_{1}^{q1-1}+C_{i2}\beta_{2}^{q1-2}+...+C_{iq1}\beta_{q1}^{0}+U_{i2}A_{iq1}+U_{i1}+\epsilon_{iq1}, \\
&Y_{iq2}=C_{i1}\beta_{1}^{q2-1}+C_{i2}\beta_{2}^{q2-2}+...+C_{iq2}\beta_{q2}^{0}+U_{i2}A_{iq2}+U_{i1}+\epsilon_{iq2}, \\
&Y_{iq3}=C_{i1}\beta_{1}^{q3-1}+C_{i2}\beta_{2}^{q3-2}+...+C_{iA_{iq3}}\beta_{A_{iq3}}^{q3-A_{iq3}}+U_{i2}A_{iq3}+U_{i1}+\epsilon_{iq3}
\end{align}

Estimation proceeds analogously to the procedure introduced in section 5.2. Step 1: In order to eliminate $U_{i1}$, first take the difference between equation 22 and 21, divided by $\bigtriangleup_{q_2q_1}A_{i}$,
\begin{align}
\frac{Y_{iq2}-Y_{iq1}}{\bigtriangleup_{q_2q_1}A_{i}}=&\frac{C_{i1}\beta_{1}^{q2-1}+C_{i2}\beta_{2}^{q2-2}...+C_{iq2}\beta_{q2}^{0}-(C_{i1}\beta_{1}^{q1-1}+C_{i2}\beta_{2}^{q1-2}...+C_{iq1}\beta_{q1}^{0})}
{\bigtriangleup_{q_2q_1}A_{i}}\\\nonumber
&+U_{i2}+\frac{\epsilon_{iq2}-\epsilon_{iq1}}{\bigtriangleup_{q_2q_1}A_{i}},
\end{align}
and then take the difference between equation 23 and 22, divided by $\bigtriangleup_{q_3q_2}A_{i}$,
\begin{align}
\frac{Y_{iq3}-Y_{iq2}}{\bigtriangleup_{q_3q_2}A_{i}}=&\frac{C_{i1}\beta_{1}^{q3-1}+C_{i2}\beta_{2}^{q3-2}+...+C_{iA_{iq3}}\beta_{A_{iq3}}^{q3-A_{iq3}}-(C_{i1}\beta_{1}^{q2-1}+...+C_{iq2}\beta_{q2}^{0})}
{\bigtriangleup_{q_3q_2}A_{i}}\\\nonumber
&+U_{i2}+\frac{\epsilon_{iq3}-\epsilon_{iq2}}{\bigtriangleup_{q_3q_2}A_{i}}.
\end{align}

Step 2: In order to remove $U_{i2}$, take the difference between equations 25 and 24 and rearrange terms to combine parameters containing the same $C_{it'}$, yielding:
\begin{align}
\frac{Y_{iq3}-Y_{iq2}}{\bigtriangleup_{q_3q_2}A_{i}}-\frac{Y_{iq2}-Y_{iq1}}{\bigtriangleup_{q_2q_1}A_{i}}=&C_{i1}(\frac{\beta_{1}^{q3-1}-\beta_{1}^{q2-1}}{\bigtriangleup_{q_3q_2}A_{i}}-\frac{\beta_{1}^{q2-1}-\beta_{1}^{q1-1}}{\bigtriangleup_{q_2q_1}A_{i}})+...\\\nonumber
&+C_{iq2}(\frac{(\beta_{q2}^{q3-q2}-\beta_{q2}^0}{\bigtriangleup_{q_3q_2}A_{i}}-\frac{\beta_{q2}^{0}}{\bigtriangleup_{q_2q_1}A_{i}})\\\nonumber
& +\frac{C_{iq2+1}\beta_{q2+1}^{q3-q2-1}}{\bigtriangleup_{q_3q_2}A_{i}}+...+\frac{C_{iA_{iq3}}\beta_{A_{iq3}}^{q3-A_{iq3}}}{\bigtriangleup_{q_3q_2}A_{i}}\\\nonumber
&+\frac{\epsilon_{iq3}-\epsilon_{iq2}}{\bigtriangleup_{q_3q_2}A_{i}}-\frac{\epsilon_{iq2}-\epsilon_{iq1}}{\bigtriangleup_{q_2q_1}A_{i}}
\end{align}

From equation 26, we see that the parameters $\beta_{t'}^{gap}$ for grandparent characteristics $C_{i1}...C_{iq2}$ between the first and second test dates are not identified, because the number of parameters exceeds the number of variables. 
By contrast, the parameters $\beta_{q2+1}^{q3-q2-1}...\beta_{A_{iq3}}^{q3-A_{iq3}}$ for grandparent characteristics $C_{iq2+1}...C_{iA_{iq3}}$ between the second and third test dates are identified and can be estimated by OLS regression, because the number of parameters equals the number of variables. Steps 3 and 4 proceed unchanged from section 5.2. Hence, whether or not the analyst assumes the more restricted DGP of equation \ref{eq:dgp} or the more general DGP with gap-specific parameters, she can estimate the $LAOE_{t}(a)$, the $PAOE_t$, and their conditional versions for durations of grandparental overlap between the second and third test date, $q_2 < A_{iq_3}<q_3$.

\end{document}